\documentclass{pacificfusion}

\usepackage[numbers]{natbib}
\usepackage{xcolor}
\usepackage{appendix}
\usepackage{subcaption}
\usepackage{mwe}
\usepackage{graphicx}
\usepackage{xr}
\usepackage{marvosym}
\usepackage{xspace} % For extra space in commands like \micron
\usepackage{booktabs} % Fancy tables

\graphicspath{{assets/images/}}

\newcommand{\rem}{\text{Re}_\text{m}}
\newcommand{\lem}{\text{Le}_\text{m}}

\title{Analysis of Preheat Propagation in MagLIF-like Plasmas}

\author[1]{Fernando Garcia-Rubio}
\author[1]{Scott Davidson}
\author[1]{C. Leland Ellison}
\author[1]{Nathan B. Meezan}
\author[1]{Douglas S. Miller}
\author[1]{Nantas Nardelli}
\author[1,2]{Adam Reyes}
\author[1]{Paul F. Schmit}
\author[1]{Hardeep Sullan}

\affil[1]{Pacific Fusion, 6082 Stewart Ave, Fremont, CA 94538}
\affil[2]{Flash Center for Computational Science, Department of Physics and Astronomy, University of Rochester, Rochester, NY 14611}

\begin{abstract}
The preheating and pre-magnetization of fusion fuel are key features in Magnetized Liner Inertial Fusion (MagLIF) configurations. 
Typically, the energy of the preheat laser is deposited in a central region of the fuel and propagates outward, generating magneto-hydrodynamic structures that impact the fuel mass distribution and magnetic flux compression during the subsequent implosion. 
We present a theoretical analysis of preheat propagation in a magnetized plasma under conditions typical for MagLIF.
The analysis is based on the acoustic time scale for the propagation of pressure disturbances being much shorter than the conductive time scale for heat diffusion.
In this regime, the preheat-driven expansion induces the stratification of the fuel and magnetic field, which accumulate in a dense outer shelf bounded by the leading shock. 
We derive self-similar solutions of the model that describe the hydrodynamic profiles of the expansion, and evaluate the evolution of the magnetic field in this configuration. 
These solutions are supported by FLASH simulations of preheat propagation. 
Our analysis shows that, asymptotically in time, the regions where the magnetization of the fuel is significant tend to become localized at the interface separating the outer shelf from the inner hot core.
We assess the implications of this stratification on the magnetic flux conservation and performance of fully integrated MagLIF FLASH simulations.

\end{abstract}

\begin{document}
\maketitle

\section{Introduction}
\label{sec:introduction}

Magnetically-driven inertial confinement fusion concepts, also referred to as pulser ICF \cite{AMPS_2025,opportunities_2025}, often rely on the compression of an embedded axial magnetic field to thermally insulate the fusion fuel.
This magnetic insulation relaxes the requirements on liner implosion velocity, opening the design parameter space to longer time scale implosions \cite{lindemuth2009fundamental}. 
Operating on a time scale of $\sim$100 ns, the Magnetized Liner Inertial Fusion (MagLIF) concept \cite{slutz2010pulsed,Gomez_PRL_2014,Yager-Elorriaga_2022} explores liner-on-target configurations that combine fuel magnetization with laser-driven fuel preheating \cite{weis2021scaling}. 
Recent theoretical efforts to scale MagLIF to larger pulsed-power drivers while preserving the magneto-hydrodynamic features of the implosions have demonstrated igniting configurations for currents exceeding 45 MA \cite{SchmitRuiz2020,Ruiz_2023_Theory,Ruiz_2023_Iscaling}, establishing MagLIF as a compelling path toward achieving facility-level gain. Indeed, this and other scaling arguments are motivating Pacific Fusion's construction of a 60 MA Demonstration System anticipated to achieve facility gain \cite{AMPS_2025}. 

During a MagLIF implosion, the externally applied field is compressed by the electrically conducting fuel in the stages following preheat. 
The performance of a MagLIF design thereby depends on the quality of this axial field compression. 
Magnetic diffusion degrades the compression by dissipating magnetic field out of the fuel region.
Additionally, extended magneto-hydrodynamic (MHD) terms such as Nernst enhance this process by convecting the field toward colder, more diffusive fuel regions \cite{velikovich2015magnetic,velikovich2019nernst}. 

Over the past decade, theoretical analyses aimed at understanding magnetic flux conservation under MagLIF conditions have primarily focused on the final stage of the implosion process, where the fuel is compressed subsonically \cite{garcia2017mass,garcia2018magnetic,garcia2018pressure,garcia2018mass,farrow2022self}. 
This focus is mainly motivated by the large temperature and axial magnetic field gradients that develop following hot spot assembly. 
However, the initial preheat phase can also influence the subsequent conservation of magnetic flux. 
The hydrodynamic structures that arise during the propagation of the preheat laser energy can redistribute the magnetic field--most notably, by accumulating it in the outer layers of the fuel--which in turn affects the overall flux conservation. 
This magnetic reconfiguration during preheat is the primary focus of the present work.

The propagation of energy deposited by an external source has been theoretically investigated in the context of laser-produced plasmas \cite{barrero1977self,barrero1980transition} and combustion initiation \cite{kurdyumov2003heat,linan2003coupling}.
In particular, the present analysis is strongly influenced by the work of Kurdyumov, S\'anchez, and Liñ\'an \cite{kurdyumov2003heat}.
In that work, the authors investigated the propagation of energy in the limit where the acoustic time scale for propagation of pressure disturbances, $t_a$, is much smaller than the time scale for conductive heat propagation, $t_c$. 
In this limit, the structure of the hydrodynamic expansion features two distinct regions: a hot, conductive core where the energy is absorbed, and a colder, denser outer layer bounded by the leading shock wave.
This is also the limit under which preheat operates in MagLIF conditions. 

In this paper, we further investigate preheat energy propagation in a plasma in the $t_a\ll t_c$ limit and examine the evolution and impact of a background axial magnetic field. 
We complete the analysis made in Ref. \cite{kurdyumov2003heat} by focusing on the case identified by the authors as `instantaneous heat deposition', characterized by a  laser deposition time $t_d$ much shorter than the acoustic time $t_a$. 
In this case, the pressures and temperatures within the propagation structure are significantly higher than those of the background fluid, enabling a re-normalization of the problem where the only input parameters are the background density, energy deposited per unit length, deposition time, and initial magnetic field intensity. 
We have then assessed the adequacy of this model in qualitatively describing the structures observed during preheat in 1D integrated MagLIF simulations with FLASH, and used it to understand how preheat dynamics impact the conservation of magnetic flux.

This paper is organized as follows. 
In Sec. \ref{sec:model}, we present the mathematical model and normalize the governing equations.
The flow structure in the asymptotic limit $t_a\ll t_c$ is discussed in Sec. \ref{sec:flow structure}.
In Sec. \ref{sec:self-similar solution}, we derive self-similar solutions for the case of constant energy deposition rate. 
The magnetic field dynamics are also discussed in this section, as well as the existence of self-similarity for more general deposition laws.
In Sec. \ref{sec:maglif conditions}, we present FLASH simulations of preheat propagation scenarios with input parameters representative of typical MagLIF conditions.
This serves two purposes: to benchmark the self-similar solutions with the FLASH code, and to examine the suitability of this analysis for MagLIF.
In Sec. \ref{sec:maglif}, we investigate the hydrodynamic structures appearing during the preheat phase of 1D integrated MagLIF simulations, and we evaluate their impact on flux conservation.
Finally, in Sec. \ref{sec:conclusions}, conclusions are drawn.

\section{Mathematical model}
\label{sec:model}

We consider the propagation of energy deposited by an external source in a magnetized plasma in cylindrical geometry. 
The background plasma is composed of equimolar deuterium-tritium (DT) gas and obeys a gamma-law equation of state with adiabatic ratio $\gamma$. 
We assume it to be fully ionized with transport coefficients given by Braginskii \cite{Braginskii1965}.
For simplicity, we neglect radiative effects and assume an instantaneous ion-electron thermal equilibration rate.

The external source, located at $r = 0$, deposits an amount of energy per unit length $E$ at a temporal rate $q(t)$ during a deposition time $t_d$.
Letting $\phi(r, t)$ denote the energy deposited per unit volume per unit time, we have $q(t) = \int_0^\infty 2\pi \phi(r,t) r \text{d}r$, and similarly $E = \int_0^{t_d} q(t)\text{d}t$. 
We consider the energy deposition rate to be sufficiently strong that the resulting pressures are significantly larger than that of the background plasma, which is otherwise at rest with a density $\rho_0$ and axial magnetic field $B_0$. 
The background magnetic field can be strong enough to modify the transport coefficients but we always assume its pressure to remain negligible (large thermal-to-magnetic pressure ratio $\beta$).

Under these hypotheses, the equations governing the propagation of the energy correspond to the conservation of mass, momentum and internal energy, and induction, reading:

\begin{equation}
    \frac{\partial \rho}{\partial t} + \dfrac{1}{r}\frac{\partial }{\partial r}\left(r \rho v\right) = 0,
    \label{eq:mass}
\end{equation}

\begin{equation}
    \rho\left(\frac{\partial v}{\partial t} + v \frac{\partial v}{\partial r} \right) = - \frac{\partial p}{\partial r},
    \label{eq:momentum}
\end{equation}

\begin{equation}
    \frac{1}{\gamma - 1}\frac{\partial p}{\partial t} + \frac{\gamma}{\gamma - 1} \dfrac{1}{r}\frac{\partial }{\partial r}\left(r v p\right) - v \frac{\partial p}{\partial r} = \dfrac{1}{r}\frac{\partial }{\partial r}\left(r \chi \dfrac{\partial T}{\partial r}\right) + \phi,
    \label{eq:energy}
\end{equation}

\begin{equation}
    \frac{\partial B}{\partial t} + \dfrac{1}{r}\frac{\partial }{\partial r}\left(r v B\right) = - \dfrac{1}{r}\frac{\partial }{\partial r}\left(r v_N B\right) + \dfrac{1}{r}\frac{\partial }{\partial r}\left(r D_m \dfrac{\partial B}{\partial r}\right).
    \label{eq:induction}
\end{equation}
Here, $\rho$, $v$, $p$, $T$, and $B$ refer to the mass density of the plasma, ion radial velocity, total (ion + electron) pressure, temperature (in energy units and assumed equal for ions and electrons), and axial magnetic field intensity, respectively.
These equations are completed with the equation of state of an ideal gas,
\begin{equation}
    p = \frac{Z + 1}{Am_p}\rho T,
\end{equation}
where $m_p$ is the proton mass and $Z=1$, $A=2.5$ are the the ionic charge and mass number of the fully ionized DT plasma. 

The transport terms $\chi$, $D_m$, and $v_N$ correspond to thermal conductivity, magnetic diffusivity, and Nernst velocity.
The latter is the only electrothermal term relevant in a large $\beta$ plasma in this geometry, and accounts for magnetic field compression and convection down the gradients of electron temperature. 
In the theoretical derivation that follows, a low magnetization limit is considered. 
This limit corresponds to a regime where externally applied magnetic fields are too weak to alter the transport properties in the perpendicular directions, and the induction equation becomes uncoupled.
In contrast, the numerical results presented later explore arbitrary magnetization levels, where transport properties are influenced by the background magnetic field.
The extent of this modification is given by the electron Hall parameter, $x_e = \omega_e\tau_e$, where $\omega_e =eB/m_ec$ is the electron cyclotron frequency. 
In the $x_e\ll1$ limit, these coefficients read 

\begin{equation}
    \chi = \gamma _0 \frac{n_e T \tau _e }{m_e}, \quad D_m = \alpha _0 \frac{c^2m_e}{4\pi e^2 n_e \tau_e}, \quad v_N = -\frac{\beta _0^{\prime\prime}}{\delta_0} \frac{\tau _e}{m_e}\frac{\partial T}{\partial r},
\end{equation}
respectively. 
In the expressions above, $n_e = Z\rho/Am_p$ refers to the electron number density, $\tau_e$ corresponds to the electron collision time, and $m_e$ is the mass of the electron. 
Notice that, in a fully ionized plasma, $\tau_e \sim T^{3/2}/Zn_e$.
For the particular case of a hydrogenic plasma, we have $\gamma_0 = 3.1616$, $\alpha_0 = 0.5129$, $\beta_0^{\prime\prime} = 3.053$, and $\delta_0 =3.7703$.
We have neglected ion thermal conductivity in this low magnetization limit as it is smaller by an order of $(m_e/m_i)^{1/2}$, with $m_i=Am_p$ being the ion mass.
We notice that, in the opposite strong magnetization limit $(x_e\gg 1)$, both the electron thermal conductivity and the Nernst velocity decrease as $1/x_e^2$, while the diffusivity increases by approximately a factor of 2. 

We assume that the external source is concentrated, meaning that its radius is smaller than the characteristic length scale of the hydrodynamic motion under consideration. 
This is a reasonable assumption when studying preheat propagation in MagLIF configurations, where the ratio of the liner's inner surface to the laser spot is roughly 3 (some examples are provided in Table \ref{tab:maglif input paramteres}). 
Consequently, the energy deposited on-axis must be transferred to the background plasma via thermal conduction.
We can then drop the energy deposition term in Eq. \eqref{eq:energy} as $\phi \sim \delta(r)$, and absorb it in the boundary conditions that complete the system of Eqs. \eqref{eq:mass}--\eqref{eq:induction}, which read 

\begin{equation}
\begin{aligned}
    r = 0:&\quad 2\pi r\chi \frac{\partial T}{\partial r} = -q, \quad \frac{\partial B}{\partial r} = 0.\\
    r \rightarrow \infty:& \quad p = v = 0, \quad \rho = \rho_0.
\end{aligned}
\label{eq:bcs}
\end{equation}

\subsection{Normalization}
\label{subsec:normalization}

The parameters $E$, $\rho_0$ and $t_d$ span a sufficient number of dimensions to construct any dimensional quantity based on mass, length, and time.
Particularly, a reference pressure can be derived as

\begin{equation}
    p_0 = \left(\frac{\gamma -1}{\gamma\pi}\right)^{1/2}\frac{\left(\rho_0 E\right)^{1/2}}{t_d}.
    \label{eq:pref}
\end{equation}
%which corresponds to the level of pressure to which the background cold plasma is expected to raise. 
Based on this, the characteristic size $r_h$ of the hot column of plasma that is expected to reach that internal energy level is

\begin{equation}
    r_h = \left(\frac{\gamma - 1}{\gamma \pi}\frac{E}{p_0}\right)^{1/2}.
    \label{eq:rhref}
\end{equation}
A characteristic temperature can similarly be defined, $T_0 = p_0m_i/(Z+1)\rho_0$.

In the absence of radiation, the energy deposited can propagate either as pressure disturbances or through heat diffusion.
In the same spirit as in the analysis performed in Ref. \cite{kurdyumov2003heat}, we define an acoustic time scale for propagation of the former as $t_a = r_h / (p_0/\rho_0)^{1/2}$. 
Similarly, heat conduction across the hot column operates in a conductive time scale $t_c = (Z+1)\gamma\rho_0 r_h^2/(\gamma-1)m_i\chi_0$, where we use the subscript $0$ to denote any transport term evaluated at $T_0$ and $\rho_0$.
The ratio between both times, $\epsilon \equiv t_a/t_c$, assesses how effectively each mechanism operates. 
We can derive

\begin{equation}
    \epsilon = \gamma_0\frac{\gamma-1}{\gamma}\frac{Z}{\left(Z+1\right)^{3/2}} \frac{m_i}{m_e}\frac{\left(T_0/m_i\right)^{1/2}\tau_{e0}}{r_h}\sim \left(\frac{m_p}{m_e}\right)^{1/2} \text{Kn},
    \label{eq:epsilon}
\end{equation}
where Kn refers to the Knudsen number of the flow, defined as the ratio between the mean free path of the ions and the characteristic length of the problem, $r_h$ in our case. 
In the case of small $\epsilon$, acoustic pressure disturbances propagate faster, and the leading edge of the flow structure becomes a shock wave.
We anticipate that this is the limit investigated in this work.
This justifies the reference pressure derived in Eq. \eqref{eq:pref}, which, by construction, implies $t_a = t_d$, ensuring that energy deposition and propagation occur on the same time scale. 
Our analysis differs in this regard from Ref. \cite{kurdyumov2003heat}, where the background gas pressure was used as the reference, leading to a characteristic acoustic time $t_a$ larger than the deposition time $t_d$.

We use $r_h$, $t_d$, $\rho_0$, $p_0$, and $B_0$ to normalize the system of Eqs. \eqref{eq:mass}--\eqref{eq:induction}. 
Keeping, for simplicity, the same nomenclature for dimensional variables and their normalized counterparts, the energy and induction equations become

\begin{equation}
    \frac{1}{\gamma}\frac{\partial p}{\partial t} + \dfrac{1}{r}\frac{\partial }{\partial r}\left(r v p\right) - \frac{\gamma - 1}{\gamma}v \frac{\partial p}{\partial r} = \dfrac{\epsilon}{r}\frac{\partial }{\partial r}\left(r T^{5/2} \dfrac{\partial T}{\partial r}\right),
    \label{eq:energy norm}
\end{equation}

\begin{equation}
    \frac{\partial B}{\partial t} + \dfrac{1}{r}\frac{\partial }{\partial r}\left(r v B\right) = \dfrac{\epsilon\beta_N}{r}\frac{\partial }{\partial r}\left(r \frac{T^{5/2}}{p} \frac{\partial T}{\partial r} B\right) + \dfrac{1}{\rem r}\frac{\partial }{\partial r}\left(\frac{r}{T^{3/2}} \dfrac{\partial B}{\partial r}\right),
    \label{eq:induction norm}
\end{equation}
while mass and momentum remain the same. 
The equation of state allows the normalized temperature to be expressed as $T = p/\rho$. 
The temperature dependence of conductivity, with a 5/2 exponent, is characteristic of electronic heat conduction.
In the induction equation above, we have introduced a redefined Nernst coefficient $\beta_N$ and the magnetic Reynolds number $\rem$ as

\begin{equation}
    \beta_N=\frac{\gamma}{\gamma-1}\frac{Z+1}{Z}\frac{\beta_0^{\prime\prime}}{\gamma_0\delta_0}, \quad \rem = \frac{r_h^2}{t_d} \frac{4\pi e^2n_{e0}\tau_{e0}}{\alpha_0c^2m_e}.
    \label{eq:betan rem}
\end{equation}

Introducing a normalized energy deposition length $f(t) = q(t)t_d/E$, the boundary condition on axis for heat flux, invoked in normalized variables, becomes

\begin{equation}
    \left. 2\epsilon rT^{5/2}\frac{\partial T}{\partial r}\right|_{r=0} = - f,
    \label{eq:heat flux bc norm}
\end{equation}
with $f(t)$ satisfying the constraint $\int_0^1f(t)\text{d}t = 1$.

\section{\texorpdfstring{Flow structure for $\epsilon \ll 1$}{Flow structure for ε << 1}}

\label{sec:flow structure}

\begin{figure}[h]
    \centering
    \includegraphics[width=0.65\linewidth]{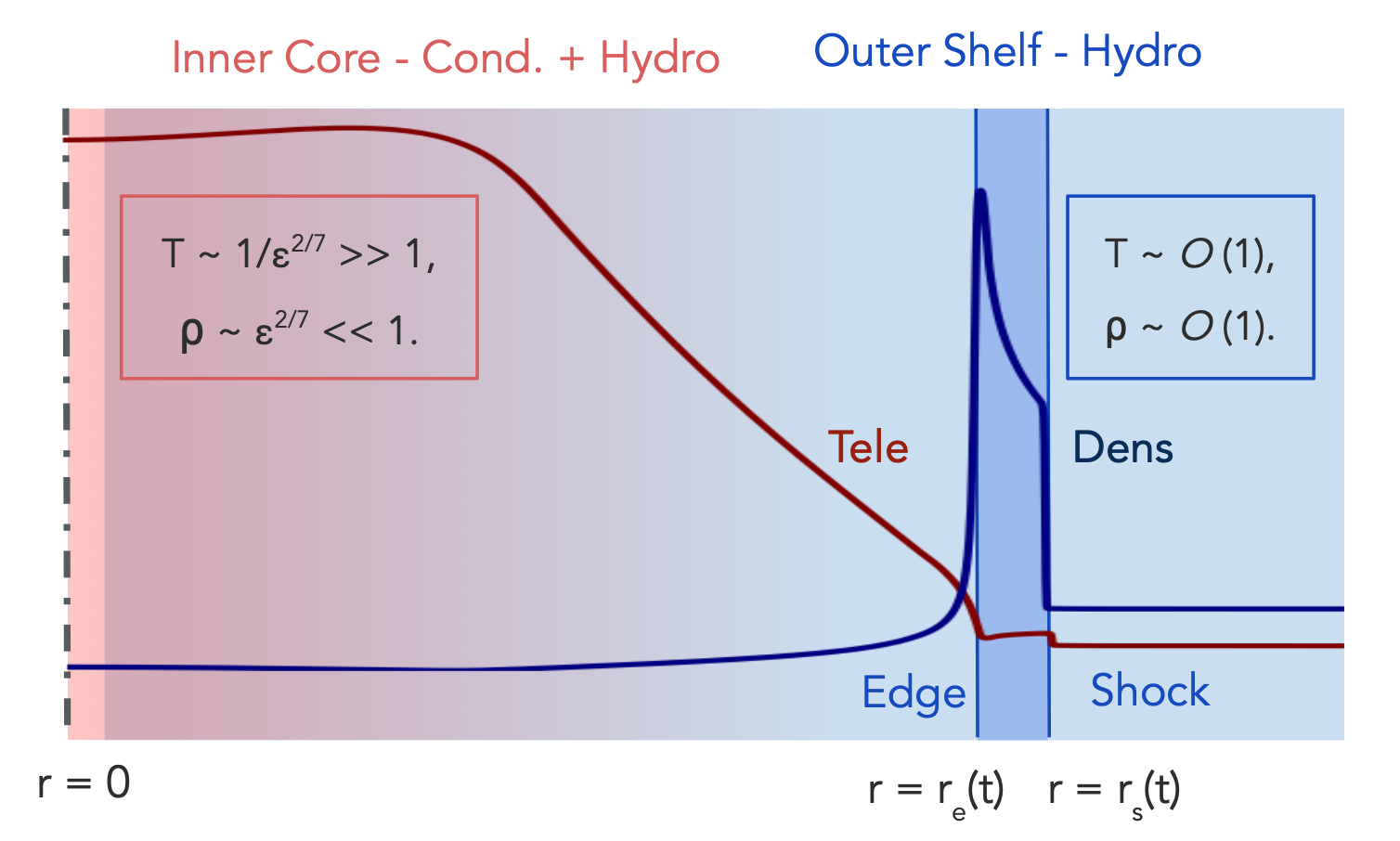}
    \caption{Sketch of the flow structure for the propagation of energy deposited by an on-axis external source (laser). The leading edge of the structure consists of a shock wave propagating radially outward. Electron temperature and plasma density profiles during preheat from a scaled 60 MA MagLIF 1D FLASH simulation are superimposed.}    
    \label{fig:sketch propagation}
\end{figure}

When the time scale for the propagation of acoustic perturbations is much shorter than the conductive time scale, $\epsilon \ll 1$, the flow structure exhibits two distinguishable regions, sketched in Fig. \ref{fig:sketch propagation}.
Thermal conduction can be neglected in the outer region of the expansion, where the normalized flow variables remain of order unity. 
This region is governed by the ideal hydrodynamic equations, with conduction representing an $O(\epsilon)$ correction. 
A strong shock placed at at $r = r_s(t)$ defines the leading edge of the expansion. 
The jump conditions across the shock yield the following values relations for the flow variables immediately behind it, denoted with the subscript $s$:

\begin{equation}
    \rho_s = \frac{\gamma + 1}{\gamma -1}, \quad v_s = \frac{2\dot{r}_s}{\gamma + 1}, \quad p_s = \frac{2\dot{r}^2_s}{\gamma + 1},
\end{equation}
where the dot denotes time derivative. 

Thermal conduction cannot be neglected everywhere, as it is relevant on-axis where the external energy is transferred to the background plasma. 
This is reflected in the boundary condition Eq. \eqref{eq:heat flux bc norm}, which enforces the ordering $T\sim O(\epsilon^{-2/7}) \gg 1 $. 
A hotter, conductive core takes therefore place filled with lower density plasma $\rho \sim O(\epsilon^{2/7})\ll 1$, assuming that pressure remains of the same order of magnitude throughout the entire structure. 
In fact, from momentum conservation one can estimate that pressure gradients in this region are small, $\partial p/\partial r\sim \rho\partial v/\partial t\sim  O(\epsilon^{2/7})\ll 1$.

This quasi-isobaric inner region is surrounded by the outer denser shelf. 
Both regions are connected through a contact interface referred to as edge, placed at $r=r_e(t)$. 
Pressure is continuous across it, which enforces a zero relative fluid velocity, $v_e = \dot{r}_e$ (fluid surface condition).
We anticipate that quasi-isobaricity allows the integration in space of energy conservation in the inner region, which using the fluid surface condition yields

\begin{equation}
    \frac{r_e^2}{\gamma}\frac{\text{d}p_e}{\text{d}t} + p_e\frac{\text{d}r_e^2}{\text{d}t} = f,
    \label{eq:energy balance}
\end{equation}
where $p_e$ stands for the pressure level in the inner core.  
As noted by Kurdyumov, S\'anchez, and Liñ\'an \cite{kurdyumov2003heat}, the relation above describes how the source energy is used to increase the internal energy of the conductive region and to displace the outer cold shelf (first and terms on the left-hand side, respectively).
If $f(t)= 0$, this equation simplifies to $\text{d}(p_er_e^{2\gamma})/\text{d}t = 0$, indicating that the evolution of the inner core is globally isentropic after the source is turned off.

We anticipate the adequacy of this two-region structure for MagLIF implosions by superimposing density and electron temperature profiles from an integrated MagLIF 1D FLASH simulation in the preheat sketch in Fig. \ref{fig:sketch propagation}. 
This run corresponds to the scaled to 60 MA case described in Sec. \ref{sec:maglif}.
It can be seen that this picture remains relevant even with a finite laser hot spot, radiation physics, and electron-ion heat exchange.

\section{Self-similar solution for a constant source}
\label{sec:self-similar solution}

In the following, we consider external sources depositing energy at a constant rate, $f(t) = 1$.
In this case, the model admits a self-similar solution in the variable $\eta = r/t^{3/4}$.  
It follows from  the boundary conditions at the edge fluid interface that pressure decreases with time as $p(r,t) = P(\eta)/t^{1/2}$, while the flow velocity evolves as $v(r,t) = V(\eta)/t^{1/4}$. 
The outer shelf and the inner core can be solved sequentially, coupled by the energy balance Eq. \eqref{eq:energy balance}.

\subsection{Outer shelf}
\label{subsec:outer shelf}

In the outer shelf region, bounded by the edge at $\eta_e=r_e/t^{3/4}$ and the strong shock at $\eta_s=r_s/t^{3/4}$, the flow evolution is governed by the ideal gas equations. 
Establishing a density dependence in this region as $\rho(r,t) = R(\eta)$, these equations, transformed under the self-similar ansatz, read

\begin{equation}
    -\frac{3}{4}\eta\frac{\text{d}R}{\text{d}\eta} + \frac{1}{\eta}\frac{\text{d}}{\text{d}\eta}\left(\eta RV\right) = 0,
    \label{eq:mass outer self-similar}
\end{equation}

\begin{equation}
    R\left[\left(V-\frac{3}{4}\right)\frac{\text{d}V}{\text{d}\eta}-\frac{1}{4}V\right] + \frac{\text{d}P}{\text{d}\eta} = 0,
    \label{eq:momentum outer self-similar}
\end{equation}

\begin{equation}
    -\frac{1}{\gamma}\left(\frac{3}{4}\eta\frac{\text{d}P}{\text{d}\eta}+\frac{1}{2}P\right) 
    + \frac{1}{\eta}\frac{\text{d}}{\text{d}\eta}\left(\eta V P\right) -\frac{\gamma-1}{\gamma}V\frac{\text{d}P}{\text{d}\eta}=0.
    \label{eq:energy outer self-similar}
\end{equation}

The jump conditions at the shock invoked in a self-similar fashion read

\begin{equation}
    R_s=\frac{\gamma+1}{\gamma-1},\quad V_s = \frac{3\eta_s}{2(\gamma+1)}, \quad P_s=\frac{9\eta_s^2}{8(\gamma+1)}.
    \label{eq:shock self-similar}
\end{equation}

It is noticeable that the integration of Eqs. \eqref{eq:mass outer self-similar}--\eqref{eq:energy outer self-similar} departing from the boundary conditions above result in a vanishing density for $\eta=\eta_e$, where $V_e = 3\eta_e/4$. 
Kurdyumov, S\'anchez and Liñ\'an \cite{kurdyumov2003heat} provided a physical explanation for this observation, based on the conservation of entropy for fluid particles evolving in ideal conditions. 
Particles compressed by the shock at earlier times are subject to larger entropy increments, theoretically diverging for $t\rightarrow 0$ since $p\sim1/t^{1/2}$.
These particles maintain their entropy as they expand, reaching the contact interface $\eta_e$ for finite times. 
Consequently, for entropy to be infinite in this environment with bounded pressure, the density must vanish locally at $\eta_e$.
Then by virtue of momentum conservation Eq. \eqref{eq:momentum outer self-similar}, the pressure profile has a zero gradient (in $\eta$ variable) at the edge. 
Bearing this in mind, the energy balance Eq. \eqref{eq:energy balance} simplifies to 

\begin{equation}
    P_e\eta_e^2=\frac{2\gamma}{3\gamma-1}.
    \label{eq:energy balance self-similar}
\end{equation}
Solving the outer shelf region reduces to finding the $\eta_s$ value such that the integration of Eqs. \eqref{eq:mass outer self-similar}--\eqref{eq:energy outer self-similar}, starting from the boundary conditions \eqref{eq:shock self-similar}, satisfies Eq. \eqref{eq:energy balance self-similar} at the position where $V_e = 3\eta_e/4$.
The results for the edge and shock positions and pressures for different $\gamma$ values are summarized in Table \ref{tab:interfaces pressures}.
Notice that the values for $\gamma = 1.4$ agree with Ref. \cite{kurdyumov2003heat}, whereas for a highly compressible gas $(\gamma = 1.1)$  $\eta_e$ naturally approaches $\eta_s$. 

\begin{table}[h]
    \centering
    \begin{tabular}{c|c c c c}
         $\gamma$ & $\eta_e$ & $\eta _s$ & $P_e$ & $P_s$ \\ 
         \hline
         \hline
         1.1 & 1.147 & 1.193 & 0.635 & 0.762 \\
         1.4 & 1.143 & 1.306 & 0.670 & 0.799 \\
         5/3 & 1.119 & 1.366 & 0.666 & 0.787 \\
         2   & 1.098 & 1.435 & 0.663 & 0.772 \\
    \end{tabular}
    \caption{Normalized positions and pressures at the edge and at the shock for different values of adiabatic ratio $\gamma$ in the case of constant energy deposition rate.}
    \label{tab:interfaces pressures}
\end{table}

\begin{figure}[h]
    \centering
    \begin{minipage}{0.48\textwidth}
        \centering
        \includegraphics[width=\textwidth]{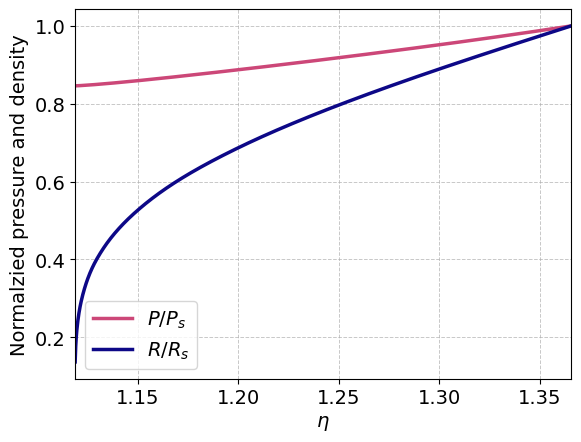}
    \end{minipage}
    \hfill
    \begin{minipage}{0.48\textwidth}
        \centering
        \includegraphics[width=\textwidth]{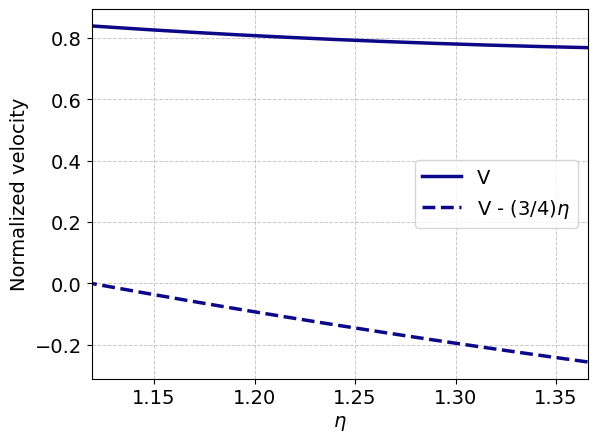}
        \end{minipage}
    \caption{Normalized pressure, density and velocity profiles in the outer shelf for an adiabatic ratio $\gamma = 5/3$ in the case of constant energy deposition rate. The domain is bounded by $\eta_e$ and $\eta_s$ values, specified in Table \ref{tab:interfaces pressures}.}
    \label{fig:profiles outer shelf}
\end{figure}

Normalized velocity, pressure and density profiles for the case $\gamma = 5/3$ are shown in Fig. \ref{fig:profiles outer shelf}.
In the right-hand panel, the fluid velocity relative to the velocity of surfaces with constant density is shown in dashed blue line.
It shows that, in this frame of reference, the fluid particles travel radially inward until they reach zero velocity at the edge.

\subsection{Inner core}
\label{subsec:inner core}

Consistently with the scaling derived in Sec. \ref{sec:flow structure}, we rescale temperature in the conductive core as $T(r,t) = \epsilon^{-2/7}\theta(r,t)$. 
Taking into account isobaricity, $p(r,t) = p_e(t)$, mass and energy conservation can be rewritten as

\begin{equation}
    \frac{\partial}{\partial t}\left(\frac{p_e}{\theta}\right) + \frac{1}{r}\frac{\partial}{\partial r}\left(r\frac{p_e v}{\theta}\right) = 0,
    \label{eq:mass inner}
\end{equation}

\begin{equation}
    \frac{1}{\gamma}\frac{\text{d}p_e}{\text{d}t} + \frac{1}{r}\frac{\partial}{\partial r}\left[r\left(p_ev-\theta^{5/2}\frac{\partial\theta}{\partial r} \right)\right] =0,
    \label{eq:ener inner}
\end{equation}
respectively. 

The on-axis boundary condition Eq. \eqref{eq:heat flux bc norm} simplifies to 

\begin{equation}
    \left. 2r\theta^{5/2}\frac{\partial \theta}{\partial r} \right|_{r=0} = -f,
    \label{eq:bc inner}
\end{equation}
which can be used to integrate Eq. \eqref{eq:ener inner}, yielding an explicit expression for the fluid velocity in the core :

\begin{equation}
    v=\frac{f}{2rp_e}-\frac{r}{2\gamma p_e}\frac{\text{d}p_e}{\text{d}t}+\frac{1}{p_e}\theta^{5/2}\frac{\partial \theta}{\partial r}.
    \label{eq:velocity expression}
\end{equation}
Inserting this expression into Eq. \eqref{eq:mass inner} results in a thermal-wave-type equation of the normalized temperature,

\begin{equation}
    \frac{\partial}{\partial t}\left( \frac{p_e}{\theta} \right) + \frac{1}{r}\frac{\partial}{\partial r}\left[ \frac{1}{\theta} \left( \frac{f}{2} -\frac{r^2}{2\gamma}\frac{\text{d}p_e}{\text{d}t} + r\theta^{5/2}\frac{\partial \theta}{\partial r}  \right)\right] = 0.
    \label{eq:thermal wave}
\end{equation}

Two additional boundary conditions are required to solve the equation above. 
The outer shelf temperature has to be retrieved at the edge which, in core rescaled variables, implies $\left.\theta\right|_{r=r_e} \rightarrow 0$.
Additionally, the cold shelf acts as a thermal insulator for the conductive core, so the heat flux must vanish at the edge. 
Using this latter boundary condition, along with null relative velocity at the edge allows one to retrieve the energy balance Eq. \eqref{eq:energy balance} by particularizing Eq. \eqref{eq:velocity expression} at $r=r_e$.

When the energy deposition rate is constant $(f = 1)$, the flow in the conductive core also evolves self-similarly, but with density and temperature following a different temporal dependence than in the outer shelf region.
The rescaled temperature can directly be assumed to behave as $\theta(r,t) = \theta(\eta)$, which leads to density evolving as $\rho(r,t)=\rho_\text{in} (\eta)/t^{1/2}$, in contrast to the profile derived in the shelf, where $\rho(r,t) = R(\eta)$. 
This fact does not invalidate the analysis, given the different orderings assumed for temperature and density in each region.

Using Eq. \eqref{eq:energy balance self-similar}, and introducing $\zeta = \eta/\eta_e$, the governing equation for temperature, Eq. \eqref{eq:thermal wave}, can be written under a self-similar transformation as

\begin{equation}
    -\frac{3}{2}\zeta\frac{\text{d}}{\text{d}\zeta}\left(\frac{1}{\theta}\right) - \frac{1}{\theta}+\frac{3\gamma-1}{\gamma\zeta}\frac{\text{d}}{\text{d}\zeta}\left[\frac{1}{\theta}\left(\frac{1}{2}+\frac{\zeta^2}{6\gamma-2}+\zeta\theta^{5/2}\frac{\text{d}\theta}{\text{d}\zeta}\right)\right] = 0,
    \label{eq:thermal wave self-similar}
\end{equation}
which is completed with the boundary conditions at the edge
\begin{align}
    \left.\theta\right|_{\zeta = 1} = 0,\quad \left.\theta^{5/2}\frac{\text{d}\theta}{\text{d}\zeta}\right|_{\zeta = 1} = 0.
    \label{eq:bc inner self-similar}
\end{align}
The mathematical structure of the thermal wave equation suggests a local expansion for the temperature profile near the edge given by $\theta = s_\theta(1-\zeta)^{2/5}$, which is consistent with the boundary conditions above. 
The eigenvalue $s_\theta$ is retrieved numerically by integrating Eq. \eqref{eq:thermal wave self-similar} toward the origin and imposing the boundary condition on axis Eq. \eqref{eq:bc inner}, assuming $f=1$.
It is computed in Table \ref{tab:eigenvalues s and sigma}, together with other related eigenvalues, for different values of the adiabatic ratio $\gamma$.

\begin{table}[h]
    \centering
    \begin{tabular}{c|c c c}
         $\gamma$ & $s_\theta$ & $s_\nu$ & $s_\sigma$ \\ 
         \hline
         \hline
         1.1 & 1.287 & 1.349 & 0.655 \\
         1.4 & 1.252 & 1.337 & 0.613 \\
         5/3 & 1.233 & 1.332 & 0.603 \\
         2   & 1.217 & 1.325 & 0.595 \\
    \end{tabular}
    \caption{Eigenvalues $s_\theta$, $s_\nu$ and normalized mass flux $s_\sigma$ at the edge for different values of adiabatic ratio $\gamma$ in the case of constant energy deposition rate.}
    \label{tab:eigenvalues s and sigma}
\end{table}

Expressing the radial velocity as $v(r,t)= 3\eta_e\nu (\zeta) /4t^{1/4}$ allows Eq. \eqref{eq:velocity expression} to be recast as

\begin{equation}
    \nu = \frac{1}{\zeta}-\frac{1-\zeta^2}{3\gamma\zeta}+\frac{6\gamma-2}{3\gamma}\theta^{5/2}\frac{\text{d}\theta}{\text{d}\zeta}.
    \label{eq:vel inner}
\end{equation}
In a similar fashion to the temperature profile, velocity can be locally expanded as $\nu = 1-s_\nu(1-\zeta)^{2/5}$ near the edge, with $s_\nu = 4(3\gamma-1)s_\theta^{7/2}/15\gamma$.

\begin{figure}
    \centering
    \includegraphics[width=0.5\linewidth]{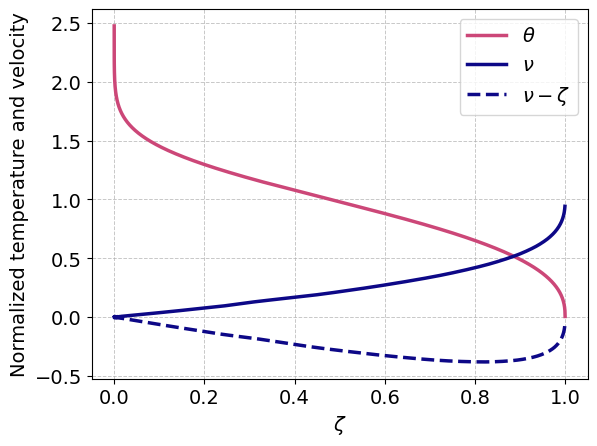}
    \caption{Normalized temperature $\theta$ and velocity $\nu$ profiles in the inner conductive core for an adiabatic ratio $\gamma = 5/3$ in the case of constant energy deposition rate.}
    \label{fig:temperature-velocity inner}
\end{figure}

The profiles for temperature and velocity are shown in Fig. \ref{fig:temperature-velocity inner} for $\gamma = 5/3$. 
Temperature diverges on axis due to the assumption of localized sources.
The relative velocity with respect to surfaces with constant temperature is depicted in
dashed blue lines. 
As in the outer shelf, the relative velocity is negative, indicating that fluid particles in the inner core come from the shelf and travel radially inward toward the origin.

To better understand the interaction between the conductive core and the shelf, it is insightful to compute the mass flux rate  per unit surface $(\dot{m})$ from the shelf into the core, which, when expressed in dimensional variables, is given by $\dot{m}=\left.\rho(\dot{r}_e-v)\right|_{r=r_e}$.
One can derive in a dimensionless fashion $\dot{m}t_d/\rho_0r_h = s_\sigma \epsilon^{2/7}/t^{3/4} $, with $s_\sigma = 2s_\theta^{5/2}/5\eta_e$ computed in Table \ref{tab:eigenvalues s and sigma}. 
From the core's perspective, the density diverges as the edge interface is approached, while the relative velocity vanishes in an inversely proportional manner.
This results in a finite mass flux, meaning the edge interface acts as an ablation front through which the core is replenished with mass coming from the compressed surrounding shelf.
This type of structures are common in ICF-type flows.
For instance, they appear at the outer interface of the shell in laser-driven capsules \cite{sanmartin1989coronal,sanz1981quasi}, or at boundary between the hot spot and the surrounding shell during capsule deceleration and stagnation \cite{sanz2005self,garnier2005multiscale}. 
The heat flux upstream of the front is what drives mass ablation and supports its structure, a process often interpreted as heat losses being recycled back via this ablated material \cite{betti2001hot}.

\subsection{Magnetic field dynamics}
\label{subsec:magnetic field dynamics}

In the unmagnetized limit investigated in this section, the evolution of the magnetic field can be determined \textit{a posteriori}, once the hydrodynamic structure of the expansion has been described.
Nernst convection is negligible in the outer shelf as it is of $O(\epsilon)$ according to induction Eq. \eqref{eq:induction norm}. 
This equation reads, after self-similar transformation,

\begin{equation}
    -\frac{3}{4}\eta\frac{\text{d}B}{\text{d}\eta} + \frac{1}{\eta}\frac{\text{d}}{\text{d}\eta}\left(\eta VB\right) = \frac{t^{1/4}}{\rem}\frac{1}{\eta}\frac{\text{d}}{\text{d}\eta}\left[ \eta \left(\frac{R}{P}\right)^{3/2}\frac{\text{d}B}{\text{d}\eta} \right].
    \label{eq:induction outer self-similar}
\end{equation}
It can be seen that self-similarity is not preserved, as the effective magnetic Reynolds number decreases with time; consequently, it is generally incorrect to assume $B(r,t) = B (\eta)$. 
We can nevertheless attempt to describe the magnetic field dynamics qualitatively.
During the interval in which the effective $\rem$ remains large, diffusion can be neglected, and the quantity $B/\rho$ is conserved along fluid particle trajectories within the shelf.
However, the degree of magnetic field compression exerted by the leading shock is case-dependent, as it separates a highly diffusive background plasma from a denser, hotter medium with a varying diffusivity. 

At the conductive core, Nernst transport becomes relevant.
According to the temperature profile depicted in Fig. \ref{fig:temperature-velocity inner}, the Nernst velocity points radially outward.
The induction equation rewritten in inner core variables becomes

\begin{equation}
    -\zeta\frac{\text{d}B}{\text{d}\zeta} + \frac{1}{\zeta}\frac{\text{d}}{\text{d}\zeta}\left(\zeta \nu B\right) = 
    \frac{2(3\gamma-1)}{3\gamma}\left[\frac{\beta_N}{\zeta}\frac{\text{d}}{\text{d}\zeta}\left(\zeta\theta^{5/2}\frac{\text{d}\theta}{\text{d}\zeta} B\right)+
    \frac{\epsilon^{3/7}P_e}{t^{1/2}\rem}\frac{1}{\zeta}\frac{\text{d}}{\text{d}\zeta}\left(\frac{\zeta}{\theta^{3/2}}\frac{\text{d}B}{\text{d}\zeta} \right)
    \right].
    \label{eq:induction inner self-similar}
\end{equation}
As in the outer shelf, the magnetic field in this region does not diffuse in a self-similar fashion. However, the effective magnetic Reynolds number is significantly higher in the conductive core by a factor $\epsilon^{-3/7}$ and, unlike in the shelf, increases over time.

In the absence of diffusion and Nernst, the magnetic field would be convected inward by the flow, expanding according to density changes as dictated by the frozen-in condition. 
When considering diffusion and Nernst, we can gain insight into how every term in Eq. \eqref{eq:induction inner self-similar} competes by expanding the expressions around the edge position.
Defining the relative inward position with respect to the edge as $\xi=1-\zeta$, we have, after some manipulation,

\begin{equation}
   \left(1-\beta_N\right)\frac{\text{d}}{\text{d}\xi}\left(\xi^{2/5}B\right) = \frac{5P_e}{2s_\theta^5}\frac{\epsilon^{3/7}}{t^{1/2}\rem}\frac{\text{d}}{\text{d}\xi}\left(\frac{1}{\xi^{3/5}}\frac{\text{d}B}{\text{d}\xi}\right).
    \label{eq:induction edge self-similar}
\end{equation}
The left-hand side of this equation represents magnetic field convection relative to the edge.
Nernst convection counteracts the fluid advection by driving the field back toward the edge. 
In the unmagnetized regime considered, the former dominates, as $1-\beta_N \approx -0.28 $ for a DT plasma.
Consequently, magnetic diffusion—the right-hand side term—remains the sole mechanism allowing magnetic field penetration from the shelf into the core.
A boundary layer develops where convection and diffusion balance each other. 
If the temporal dependence of the diffusivity coefficient is ignored, this equation reduces to that of the magnetic boundary layer arising in Nernst thermal waves, presented after Eq. (49) in Ref. \cite{garcia2017mass}, albeit with different boundary conditions.
A straightforward estimation of the terms in Eq. \eqref{eq:induction edge self-similar} yields the characteristic thickness of this layer, $\epsilon _B \sim \left(\epsilon^{3/7}/t^{1/2}\rem\right)^{1/2}$, which indicates that more magnetic field diffuses into the core at earlier times.

\subsection{A note on the existence of self-similar solutions}
\label{subsec:note}

It might seem counterintuitive that the flow structure admits a self-similar solution given the existence of three input parameters, $E$, $\rho_0$, and $t_d$, which span all relevant dimensions in the problem.
Effectively, self-similar flows typically arise in problems where characteristic dimensional scales cannot be determined from the initial parameters alone. 
This apparent paradox stems from the deposition time $t_d$ not being a true input parameter, as it merely defines the temporal limit of the solution's applicability. 
The flow structure for $t<t_d$ is not affected whatsoever by the particular value of $t_d$.
Instead, the background density and the energy deposited per unit length per unit time are the only two relevant parameters. 
The self-similar variable $r/t^{3/4}$ naturally arises from the only possible dimensional combination that can be formed which does not include mass, being length$^{4}$ / time$^3$.
Since the similarity exponent has been determined by dimensional considerations, this process is a self-similar problem of first kind \cite{zel2002physics}.

More generally, self-similar solutions also exist for power deposition laws of the form $q(t)\sim t^\alpha$, with $\alpha \geq 0$. 
In such cases, $q(t)/t^\alpha$ emerges as the relevant input parameter together with $\rho_0$. 
The self-similar variable takes the form $r/t^{(\alpha +3)/4}$, recovering the constant-deposition rate case when $\alpha = 0$. 
Pressure evolves as $p\sim t^{(\alpha-1)/2}$, while temperature at the inner core follows $T\sim t^{2\alpha/7}$.
Notice that, according to the pressure scaling, the assumption of a strong shock might not be valid for early times when $\alpha > 1$. 
This fact was already noticed by Kurdyumov, S\'anchez and Liñ\'an \cite{kurdyumov2003heat}, although the authors did not explicitly recognized self-similarity for the general case $q(t)\sim t^\alpha$.

Intermediate self-similar solutions may also emerge after a transient period following the source shutoff.
Since $p_er_e^{2\gamma}$ is conserved now, the total entropy introduced into the conductive core becomes the relevant parameter.
The motion is described in this case by the variable $r/t^{1/(\gamma + 1)}$. 
Due to the absence of energy supply, both the pressure and the temperature at the inner core decrease in time as $p\sim t^{-2\gamma/(\gamma + 1)}$, $T\sim t^{-2(3\gamma -1)/7(\gamma + 1)}$, respectively.
Notice however that the validity of this solution is subject to pressure being much larger than its ambient value and the inner core remaining  conductive.
The latter condition may break down before the first one inevitably fails, in which case the two-region description no longer applies and a Sedov-type expansion \cite{sedov1946propagation} is established. 
This transition has not been observed within the preheat time scale of MagLIF simulations reported in Sec. \ref{sec:maglif}, and its analysis is deferred to future work.

\section{Preheat propagation in MagLIF conditions}
\label{sec:maglif conditions}

We now assess the suitability of this analysis for characteristic MagLIF conditions during preheat. 
To this end, we derive numerical expressions of the key parameters that emerged from the theoretical derivation. 
The analysis formally relies on the smallness of the acoustic-to-conductive time ratio, $\epsilon$. 
When expressed in typical values for preheat energy, fuel fill density, and deposition time, we can derive

\begin{equation}
    \epsilon = 0.30 \left( \dfrac{E}{10 \text{ kJ/cm}} \right)^{3/4} \left( \dfrac{\rho_0}{1 \text{ mg/cm}^3} \right)^{-7/4} \left( \dfrac{t_d}{10 \text{ ns}} \right)^{-5/2},
    \label{eq:eps numerical}
\end{equation}
where the smallness of the coefficient indicates that acoustic disturbances are indeed expected to outpace thermal waves during preheat, leading to the formation of a strong shock that bounds the expansion.
We have assumed $\gamma = 5/3$ for all numerical expressions presented in this section.

The characteristic size of the hot column and the resulting pressures, in turn, can be expressed as

\begin{equation}
    r_h = 1.89 \left( \dfrac{E}{10 \text{ kJ/cm}} \right)^{1/4} \left( \dfrac{\rho_0}{1 \text{ mg/cm}^3} \right)^{-1/4} \left( \dfrac{t_d}{10 \text{ ns}} \right)^{1/2} \text{ mm},
    \label{eq:rh numerical}
\end{equation}

\begin{equation}
    p_0 = 356 \left( \dfrac{E}{10 \text{ kJ/cm}} \right)^{1/2} \left( \dfrac{\rho_0}{1 \text{ mg/cm}^3} \right)^{1/2} \left( \dfrac{t_d}{10 \text{ ns}} \right)^{-1} \text{ kbar},
    \label{eq:p0 numerical}
\end{equation}
respectively.
The column size is comparable to the inner liner radius considered in MagLIF. 
For reference, the baseline design in Ref. \cite{Ruiz_2023_Iscaling} for a 20 MA current drive features an inner radius of $R_\text{in} = 2.325$ mm, while the configuration scaled to a 60 MA drive has $R_\text{in} = 2.915$ mm.

Using these expressions, we can derive numerical expressions for characteristic densities and temperatures at both the shelf and the core. 
For the shelf density, we take the value immediately after the strong shock, $\rho_{\text{shelf}} = 4\rho_0$, while the analysis of the conductive core suggests $\rho_{\text{core}} =\epsilon^{2/7}\rho_0$, yielding
%\begin{equation}
%    \rho_{\text{shelf}} = 4 \left( \dfrac{\rho_0}{1 \text{ mg/cm}^3} \right) \text{ mg/cm}^3,
%    \label{eq:rho shelf numerical}
%\end{equation}
%and
\begin{equation}
    \rho_{\text{core}} = 0.71 \left( \dfrac{E}{10 \text{ kJ/cm}} \right)^{3/14} \left( \dfrac{\rho_0}{1 \text{ mg/cm}^3} \right)^{1/2} \left( \dfrac{t_d}{10 \text{ ns}} \right)^{-5/7} \text{ mg/cm}^3.
    \label{eq:rho core numerical}
\end{equation}
Characteristic temperature values follow from the application of  the equation of state considering $p_0$ as reference pressure, reading

\begin{equation}
    T_{\text{shelf}} = 116 \left( \dfrac{E}{10 \text{ kJ/cm}} \right)^{1/2} \left( \dfrac{\rho_0}{1 \text{ mg/cm}^3} \right)^{-1/2} \left( \dfrac{t_d}{10 \text{ ns}} \right)^{-1} \text{ eV},
    \label{eq:temp shelf numerical}
\end{equation}
and
\begin{equation}
    T_{\text{core}} = 657 \left( \dfrac{E}{10 \text{ kJ/cm}} \right)^{2/7} \left( \dfrac{t_d}{10 \text{ ns}} \right)^{-2/7} \text{ eV}.
    \label{eq:temp core numerical}
\end{equation}
Interestingly, the core temperature remains independent of the background density according to the model.

\subsection{Unmagnetized preheat propagation}
\label{subsec:unmagnetized propagation}

The theoretical analysis is benchmarked against FLASH simulations of preheat propagation.
FLASH is a multi-physics, adaptive mesh refinement, finite-volume Eulerian, multi-group radiation magnetohydrodynamics code \cite{Fryxell2000, Tzeferacos2015}. 
It solves the hydrodynamic description of a magnetized plasma including a three-temperature extension for electron, ion and radiation temperature fields. 
We consider a setup where an external source (laser) deposits energy in an unmagnetized background plasma at rest and at room temperature.
We have chosen as representative parameters $E = 10\text{ kJ/cm}$, $\rho_0 = 5\text{ mg/cm}^3$, and $t_d = 10\text{ ns}$, which yields $\epsilon = 0.018$.
The laser spot radius is 20 times smaller than the resulting expansion radius, $r_h = 0.126$ cm, and its energy is entirely absorbed by the electrons. 
In the simulation results described below, we have purposely increased the ion-electron equilibration frequency to ensure quick local thermodynamic equilibration. 
We have also assumed a gamma-law equation of state with $\gamma = 5/3$, and imposed a constant Coulomb logarithm value of 7.0 for the electron collision time.

\begin{figure}[h]
    \centering
    \includegraphics[width=0.85\linewidth]{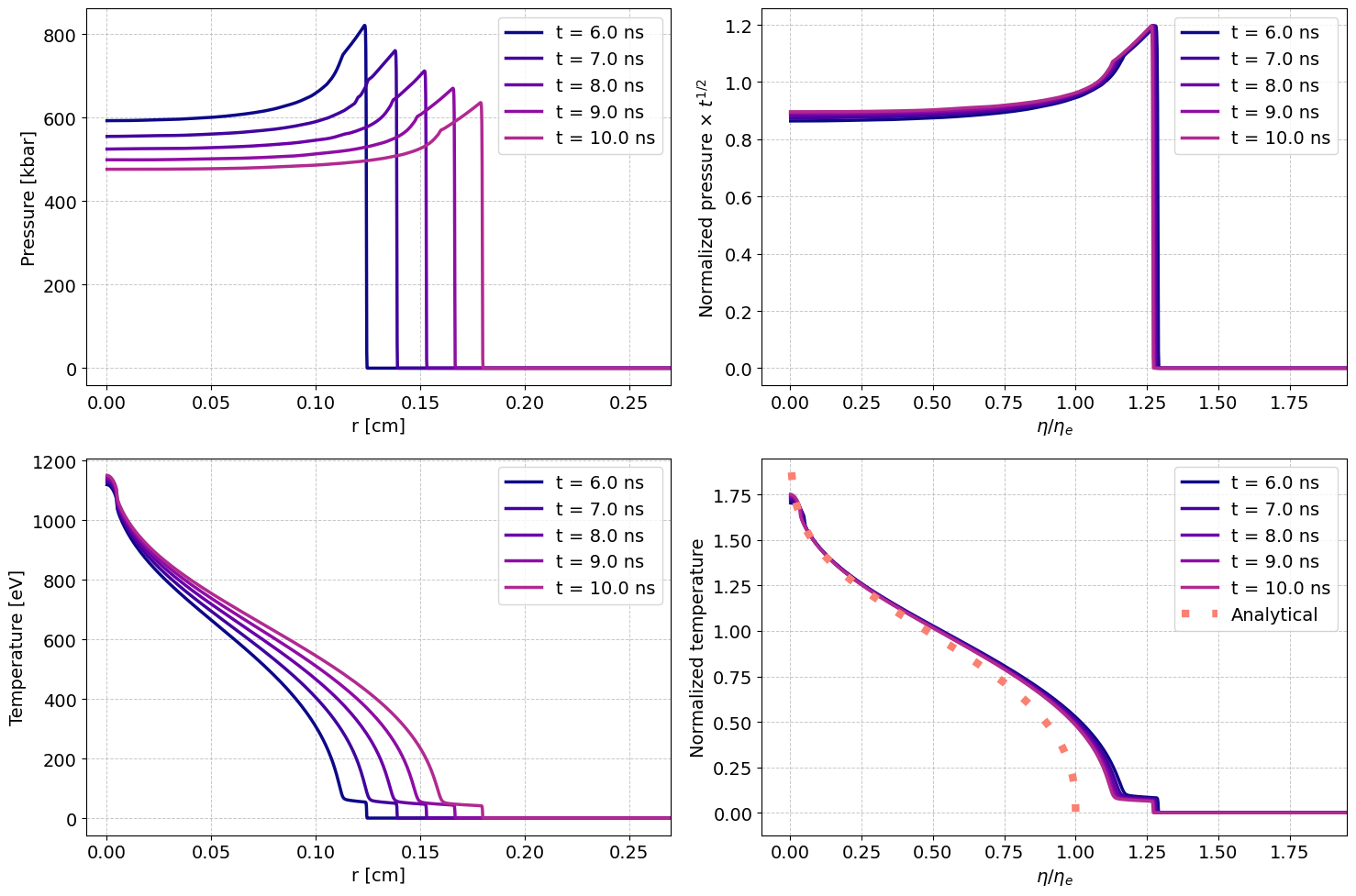}
    \caption{Pressure and temperature profiles of a FLASH preheat simulation with MagLIF-relevant input parameters $E = 10\text{ kJ/cm}$, $\rho_0 = 5\text{ mg/cm}^3$, and $t_d = 10\text{ ns}$. Each row displays the same data with the left panels plotting dimensional quantities and the right panels their normalized version under self-similar transformation. Pressure has been normalized with $P_e\times p_0$, while temperature uses $T_\text{core}$. The dotted line in the bottom-right panel corresponds to the semi-analytical temperature profile at the core, solution from Eq. \eqref{eq:thermal wave self-similar}.}
    \label{fig:preheat profiles unmagnetized maglif}
\end{figure}

Pressure and temperature profiles for MagLIF-relevant conditions are shown in Fig. \ref{fig:preheat profiles unmagnetized maglif}.
It can be seen that self-similarity is well maintained during the second-half of the laser pulse, as evidenced by the collapse of the profiles when the self-similar transformation is applied.
The temperature profile reveals the two distinctive regions identified by the analysis: a hot core surrounded by cold shelf. 
The expression for the core temperature, Eq. \ref{eq:temp core numerical}, seems appropriate, as evidenced by the normalized profile reaching unity. 
However, the pressure in the conductive core region is not perfectly flat. 
We attribute this deviation to the finite value of $\epsilon$ under such conditions. 
As derived in Sec. \ref{sec:flow structure}, normalized pressures gradients in the core scale as $\epsilon ^{2/7}\sim  0.32$, which is small but not negligibly small.
As a result, the position of the shock and the edge are ahead of what the theory predicts, and the core temperature profile deviates from the semi-analytical solution derived from Eq. \eqref{eq:thermal wave self-similar}.

\begin{figure}[h]
    \centering
    \includegraphics[width=0.85\linewidth]{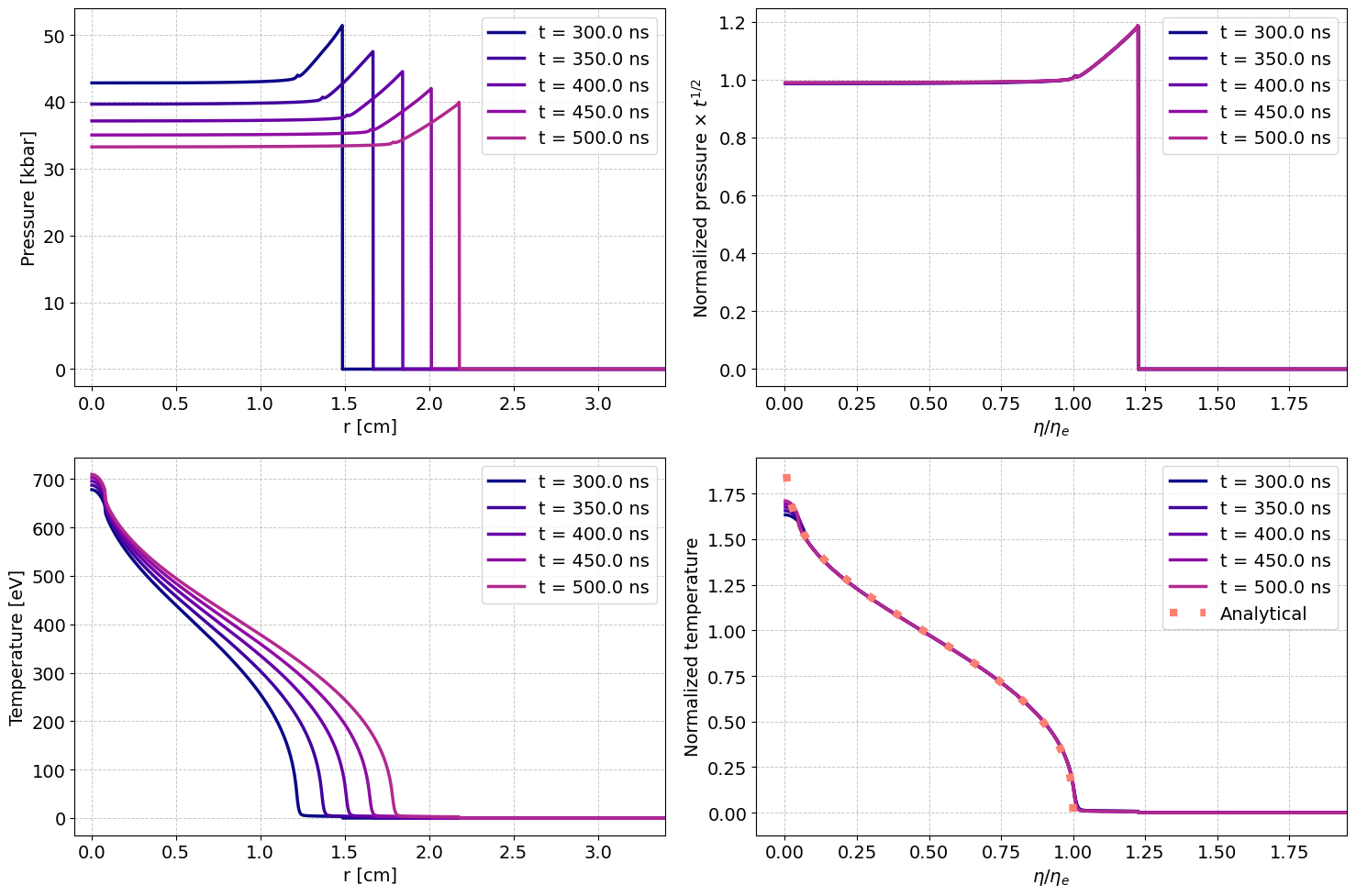}
    \caption{Pressure and temperature profiles of a FLASH preheat simulation with input parameters that ensure a small $\epsilon \sim 10^{-6}$: $E = 100\text{ kJ/cm}$, $\rho_0 = 5\text{ mg/cm}^3$, and $t_d = 500\text{ ns}$. Each row displays the same data with the left panels plotting dimensional quantities and the right panels their normalized version under self-similar transformation. Pressure has been normalized with $P_e\times p_0$, while temperature uses $T_\text{core}$. The dotted line in the bottom-right panel corresponds to the semi-analytical temperature profile at the core, solution from Eq. \eqref{eq:thermal wave self-similar}.}
    \label{fig:preheat profiles unmagnetized eps}
\end{figure}
%TODO: check coefficient in temperature

Different input conditions yielding a smaller $\epsilon=5.6\times10^{-6}$ are explored in Fig. \ref{fig:preheat profiles unmagnetized eps}. 
This case shows excellent agreement with the theory.
The pressure profile is flat in the core, which is bounded by the position predicted by theory $\eta =\eta_e$, and the temperature profile agrees well with the semi-analytical solution.
Although not shown in this figure, we remark that self-similarity is fully established in this low $\epsilon$ case after the first 20\% of the laser pulse.

\begin{figure}[h]
    \centering
    \includegraphics[width=0.85\linewidth]{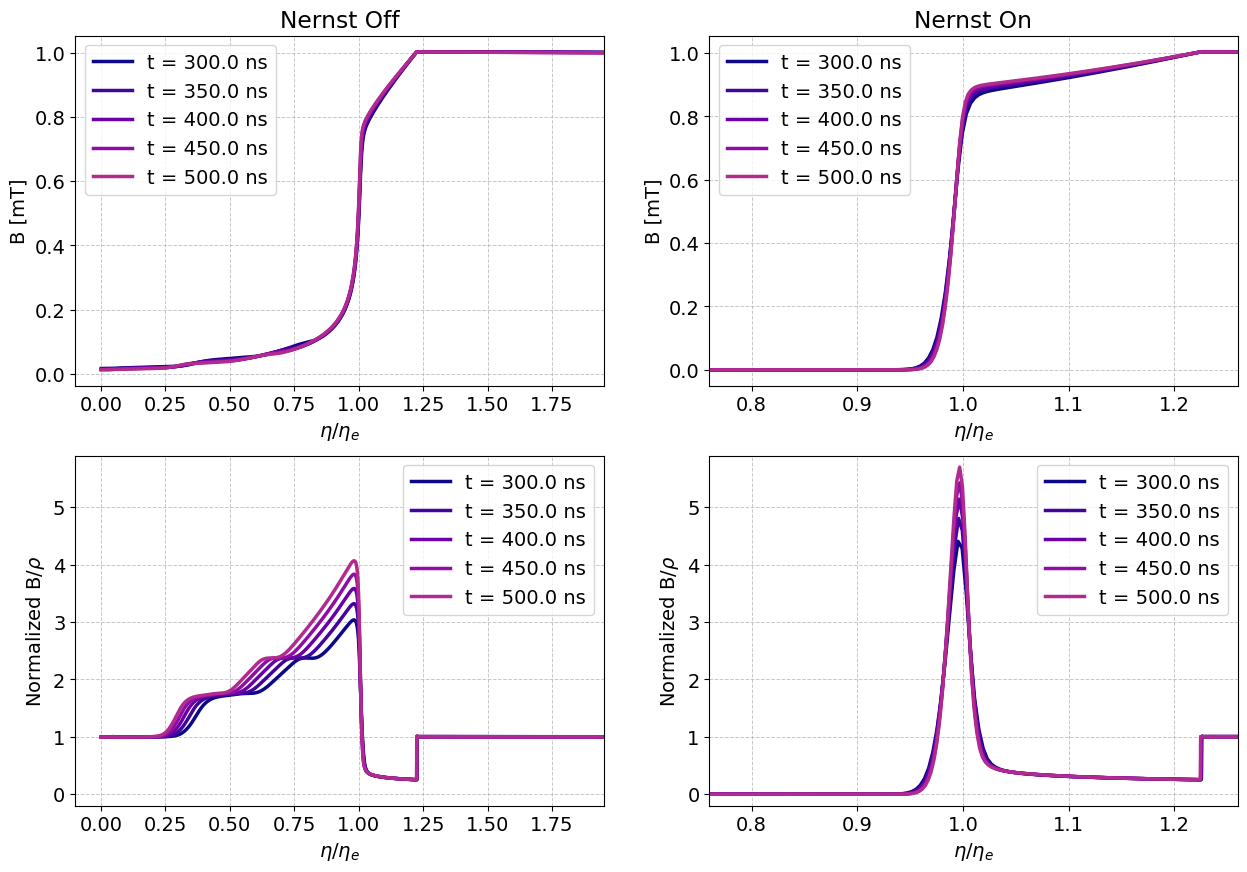}
    \caption{Self-similar profiles of axial magnetic field $B$ and ratio between magnetic field and density $B/\rho$ (normalized to its initial value) in FLASH preheat simulations with input parameters $E = 100\text{ kJ/cm}$, $\rho_0 = 5\text{ mg/cm}^3$, $t_d = 500\text{ ns}$, and $B_0 = 10^{-3}\text{ T}$. The Nernst term is disabled in the simulation results displayed on the left panels and enabled in those on the right.}
    \label{fig:magnetic field unmagnetized}
\end{figure}

An initial axial magnetic field of 1 mT is applied in the small-$\epsilon$ simulation setup. 
Its strength is low enough to avoid influencing the hydrodynamic profiles of the expansion. 
The field evolution is depicted in Fig. \ref{fig:magnetic field unmagnetized}, comparing cases with the Nernst term disabled (left panels) and enabled (right panels).
The dynamics are governed by the magnetic Reynolds number in the outer shelf, as defined in Eq. \eqref{eq:betan rem}.
Based on the input parameters, it can be expressed as

\begin{equation}
    \rem = 1.22\times 10^4 \left( \dfrac{E}{10 \text{ kJ/cm}} \right)^{5/4} \left( \dfrac{\rho_0}{1 \text{ mg/cm}^3} \right)^{-5/4} \left( \dfrac{t_d}{10 \text{ ns}} \right)^{-3/2}.
    \label{eq:rem numerical}
\end{equation}

In the particular case considered for Fig. \ref{fig:magnetic field unmagnetized}, we have a moderate value  $\rem = 82$.
Consequently, the magnetic field is barely compressed by the leading shock, which propagates in an otherwise highly diffusive medium. 
Its evolution in the outer shelf is unaffected by the Nernst term, consistent with the theoretical model. 
The quantity $B/\rho$ is plotted in the bottom panels. 
We recall that in the ideal limit, where resistive and Nernst effects are negligible, $B/\rho$ is conserved along fluid trajectories.
While this conservation does not hold across the leading shock, this quantity remains approximately flat in the outer shelf region, which indicates that diffusion effects are negligible here in spite of the moderate $\rem$ value.
In contrast, the time-increasing region immediately after the shelf reveals that fluid particles expanding into the core cannot conduct the necessary current for the magnetic field to evolve in ideal conditions.  
This supports the theoretical argument made before that diffusion is the main mechanism by which the magnetic field penetrates into the conductive core.

%TODO: check Nernst term here

Nernst convection narrows the boundary layer around the edge interface where diffusion is important. 
Notice that the Nernst-on panels are zoomed-in around this interface. 
It opposes the penetration of magnetic field from the shelf, sustaining a fully unmagnetized plasma column in the central part of the core. 
It is important to remark that even without Nernst, the hydrodynamic structure of the flow still leads to a strong demagnetization of the inner core asymptotically in time--an effect further amplified when Nernst convection is present.  

\subsection{Magnetized preheat propagation}
\label{subsec:magnetized propagation}

We analyze now preheat propagation in relevant MagLIF conditions by considering the initial representative parameters $E = 10\text{ kJ/cm}$, $\rho_0 = 5\text{ mg/cm}^3$, and $t_d = 10\text{ ns}$, together with an initial axial magnetic field of 30 T.
In this case, the magnetic field intensity is enough to modify the hydrodynamic motion. 
This coupling is characterized by the electron Hall parameter.
We can estimate it in the core region using the reference values for density and temperature in Eqs. \eqref{eq:rho core numerical} and \eqref{eq:temp core numerical}, respectively, and taking a $B_\text{core} =\epsilon^{2/7}B_0$,
\begin{equation}
    \left.\omega _e \tau _e\right|_{\text{core}} = 6.09 \left( \dfrac{E}{10 \text{ kJ/cm}} \right)^{3/7} \left( \dfrac{\rho_0}{1 \text{ mg/cm}^3} \right)^{-1} \left( \dfrac{t_d}{10 \text{ ns}} \right)^{3/7}  \dfrac{B_0}{10 \text{ T}}.
    \label{eq:ehall numerical}
\end{equation}
When evaluated with the simulation conditions, it yields $ \left.\omega _e \tau _e\right|_{\text{core}} = 3.65$.

\begin{figure}[h]
    \centering
    \includegraphics[width=0.85\linewidth]{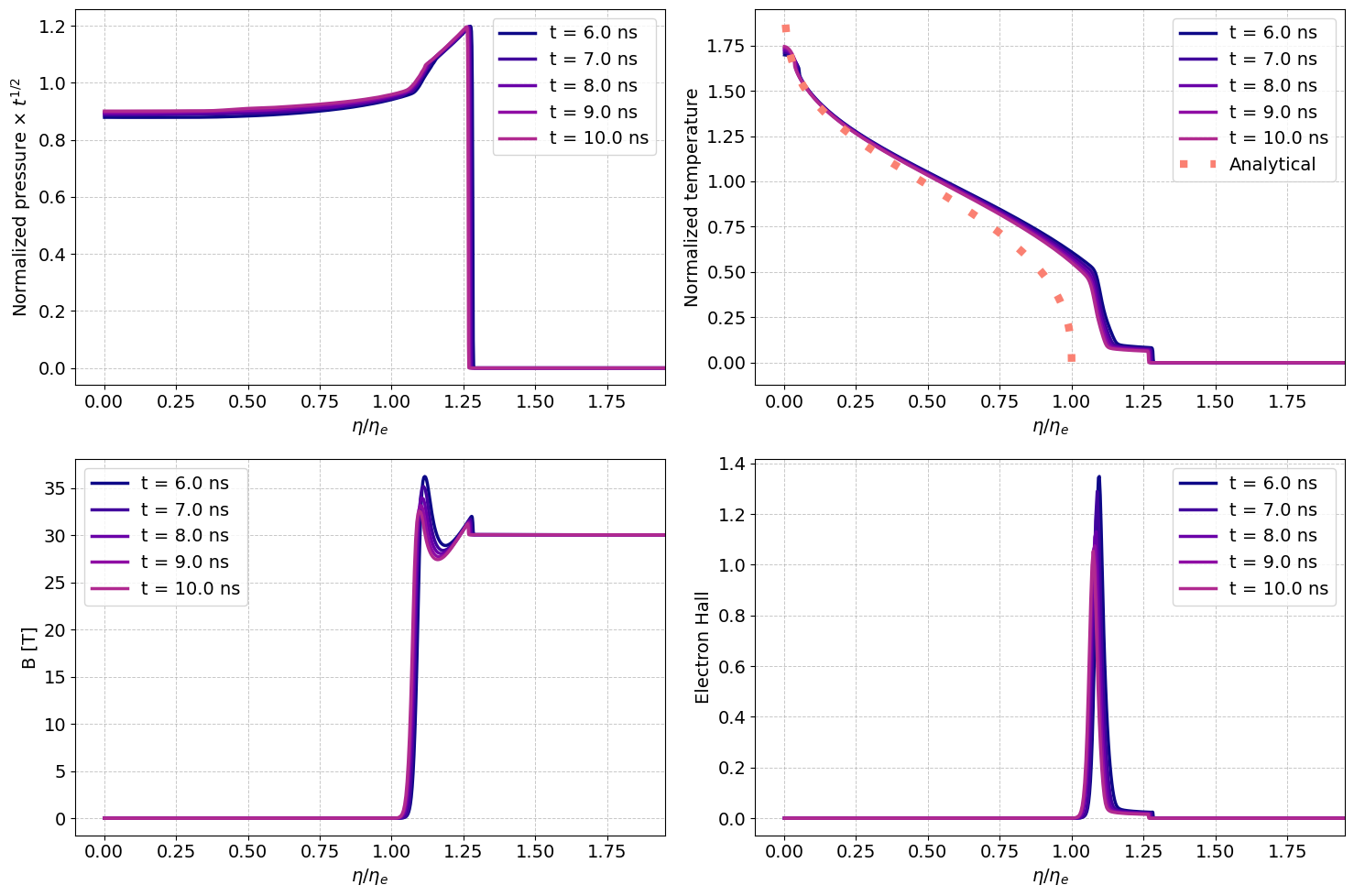}
    \caption{Self-similar profiles of normalized pressure, temperature, axial magnetic field $B$, and electron Hall parameter in a FLASH preheat simulation with MagLIF-relevant input parameters $E = 10\text{ kJ/cm}$, $\rho_0 = 5\text{ mg/cm}^3$, $t_d = 10\text{ ns}$, and $B_0 = 30\text{ T}$. The dotted line in the top-right panel corresponds to the semi-analytical temperature profile at the core, solution from Eq. \eqref{eq:thermal wave self-similar}.}
    \label{fig:preheat profiles magnetized}
\end{figure}

Profiles for pressure, temperature, magnetic field, and electron Hall parameter are shown in Fig. \ref{fig:preheat profiles magnetized}.
In spite of the coupling with the induction equation, self-similarity is roughly maintained by the hydrodynamic quantities towards the end of the laser pulse. 
In particular, the evolution of pressure is barely affected, closely resembling the unmagnetized case shown in Fig. \ref{fig:preheat profiles unmagnetized maglif}.
As a consequence of thermal insulation, the conductive core remains hotter and presents a boundary layer with sharp temperature gradients behind the edge interface. 
The rate at which mass is ablated into the core is therefore reduced as pressure is roughly maintained.

These MagLIF-relevant input conditions lead to a larger magnetic Reynolds number in the outer shelf, $\rem = 1.63\times10^3$. 
The leading shock is able to compress the magnetic field, although to a lesser extent than the accompanying density jump. 
This compression weakens over time, consistent with the theoretical analysis in Subsec. \ref{subsec:magnetic field dynamics}, which identified a time-decreasing effective magnetic Reynolds number.
The magnetic boundary layer that forms behind the edge interface exhibits a peak value of enhanced field compression due to Nernst convection.
This distinctive feature was also observed in the analysis of Nernst thermal waves presented in Ref. \cite{garcia2017mass}.
The magnetic field arrangement results in an electron Hall profile that peaks in a narrow region immediately behind the edge interface, locally insulating the core region, whose central part remains otherwise unmagnetized.
The peak electron Hall value decreases over time as a consequence of the weaker shock compression of the magnetic field.
It is predicted by Eq. \eqref{eq:ehall numerical} within a factor of $\sim2$, despite being derived under unmagnetized assumptions.
It is worth noting that the maxima in the magnetic field and electron Hall profiles are not aligned, with the electron Hall peaking at a slightly smaller radius.

\section{Implications for integrated MagLIF implosions}
\label{sec:maglif}

In this section, we investigate the impact of preheat dynamics on integrated MagLIF simulations involving pre-magnetized, preheated DT fuel inside beryllium liners.
We replicate the MagLIF current scaling study of Ref. \cite{Ruiz_2023_Iscaling}, which examines a baseline configuration of a target driven at a reference current of 20 MA, and scales the target parameters in order to maintain several dimensionless quantities at all levels of current. 
This scaling method ensures similar target evolution for all cases of coupled drive current, including the liner response to the current drive, the adiabatic heating and stagnation mechanics of the compressed DT fuel, and the impact of magnetohydrodynamic instabilities on the liner integrity, among others.
For currents exceeding 45 MA in the study, the fuel undergoes significant self-heating, exceeding a generalized Lawson ignition parameter \cite{Knapp_2022} $\chi = 1$. 
We focus accordingly on one-dimensional (1D) FLASH simulations of two configurations: the sub-ignited baseline 20 MA case, and the igniting scaled 60 MA case relevant to the Pacific Fusion's Demonstration System \cite{AMPS_2025}.
We aim to quantify the impact of Nernst convection on the performance scaling of MagLIF by turning that term on and off in the FLASH model.
Details on the FLASH simulation setup and performance metrics are thoroughly provided in Section 3.6 in Ref. \cite{Flash_Validation_2025}. 
We summarize the relevant input parameters in Table \ref{tab:maglif input paramteres}.
Based on them, the value of $\epsilon$ as provided by Eq. \eqref{eq:eps numerical} is 0.023 and 0.042 for the baseline and scaled configurations, respectively, which highlights the adequacy of our theoretical model to understand the preheat physics.

\begin{table}[h]
    \centering
    \begin{tabular}{c|cccccc}
          & $R_d$ [mm] & $R_i$ [mm] & $\rho_0$ [mg/cm$^3$]  & $E$ [kJ/cm] & $t_d$ [ns] &  $B_0 [T] $ \\ 
         \hline
         \hline
         Baseline 20 MA & 0.75 & 2.32 & 2.25 & 2.10  & 10 & 14\\
         Scaled 60 MA & 0.94 & 2.91 & 4.02 & 18.88 & 10 & 28.5
    \end{tabular}
    \caption{Input parameters of the baseline 20 MA and scaled 60 MA MagLIF configurations: laser deposition radius $R_d$, initial inner liner position $R_i$, initial fuel density $\rho_0$, preheat energy per unit length $E$, laser deposition time $t_d$, and initial axial magnetic field strength $B_0$. The preheat per unit length has been derived taking an imploding length of 1.0 cm for the baseline design and 1.78 cm for the scaled configuration.}
    \label{tab:maglif input paramteres}
\end{table}

\begin{figure}[h]
    \centering
    \includegraphics[width=0.85\linewidth]{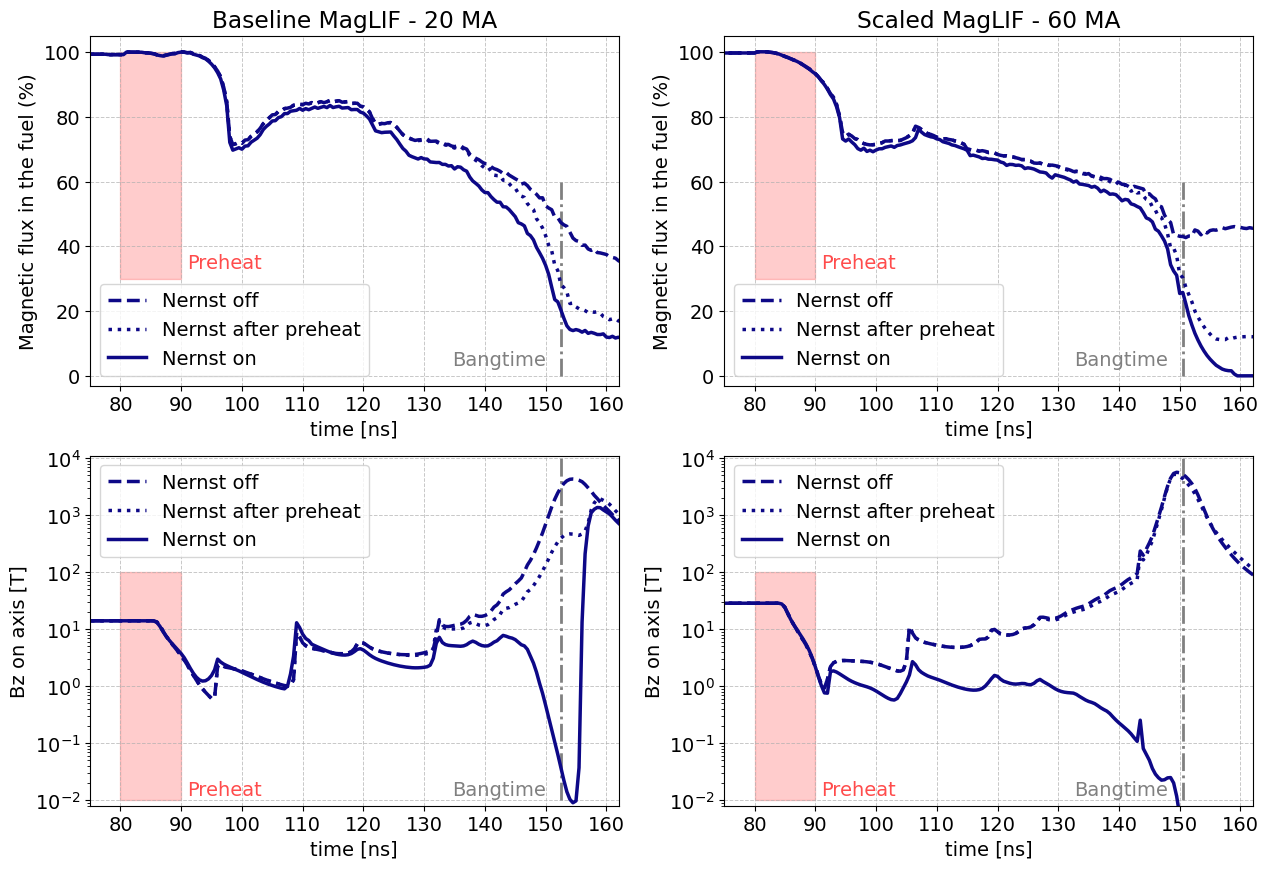}
    \caption{Evolution of the magnetic flux in the fuel (top) and the on-axis magnetic field value (bottom) for 1D MagLIF FLASH simulations with 20 MA current drive (left) and 60 MA current drive (right). In the run labeled `Nernst after preheat', Nernst convection is activated at t = 126 ns.}
    \label{fig:maglif magnetic flux losses}
\end{figure}

We start by analyzing the conservation of magnetic flux in the fuel for each configuration, shown in Fig. \ref{fig:maglif magnetic flux losses}. 
In the absence of magnetic diffusion and extended MHD terms, the axial magnetic field in the fuel is ideally compressed throughout the implosion, resulting in flux conservation \cite{velikovich1985hydrodynamics}.
As seen in Fig. \ref{fig:maglif magnetic flux losses}, there are two primary events contributing to the degradation of this compression. 
One occurs immediately after the preheat phase (which takes place from 80 ns - 90 ns), when the expanding wave in the fuel encounters the liner. 
In this first event, approximately $30\%$ of the magnetic flux is lost in both configurations regardless of the presence of Nernst convection. 
The second event happens gradually during the subsonic compression of the fuel, where the Nernst term has a significant impact. 
To gain more confidence in flux losses during this second stage, we have verified that FLASH reproduces the structure of Nernst thermal waves developing in isobaric conditions, as detailed in Appendix \ref{sec:nernst thermal wave}.

By bangtime, $50\%$ ($60\%$) of the flux is lost to the liner without Nernst in the baseline (scaled) configuration, while the inclusion of Nernst enhances this loss to $80\%$ ($75\%$). 
For reference, the seminal work by Slutz \textit{et al.} \cite{slutz2010pulsed} reports that $75\%$ of the flux is lost from the fuel in 20 MA-class targets with Nernst--despite differing liner, fuel, magnetic field, and preheat parameters from those considered here--whereas a milder $25\%$ is lost in its absence.
More recently, Lewis \textit{et al.} \cite{Lewis_2021} inferred that the ratio of magnetic flux at peak burn between Nernst-off and Nernst-on runs ranged from 1.25 to 5.5 for 16 MA current-driven beryllium liners filled with DD fuel, as the preheat energy was varied from 0.5 kJ to 2 kJ.

The temporal evolution of the on-axis magnetic field $B_z$, also shown in Fig. \ref{fig:maglif magnetic flux losses}, reveals a distinct field arrangement in the fuel when Nernst convection is included. 
Nernst causes significant demagnetization at the center of the fuel, consistent with predictions from our theoretical model. 
Although a sequence of waves reflecting between the inner liner surface and the axis advects some of the magnetic field back inward, central demagnetization persists after isobaricity is reestablished. 
This behavior contrasts with the case without Nernst convection, where the magnetic field on axis is subsonically compressed.

To elucidate the role of Nernst in the two distinct phases of magnetic flux loss, we designed a simulation where Nernst was activated after preheat once isobaricity was restored.
This roughly occurs at 126 ns, when the convergence ratio reaches 1.55 in the baseline configuration and 1.71 for the scaled case. 
As shown in Fig. \ref{fig:maglif magnetic flux losses}, omitting Nernst during preheat fails to capture the central demagnetization and underestimates the total flux loss by approximately $10\%$.

\begin{figure}[h]
    \centering
    \includegraphics[width=0.85\linewidth]{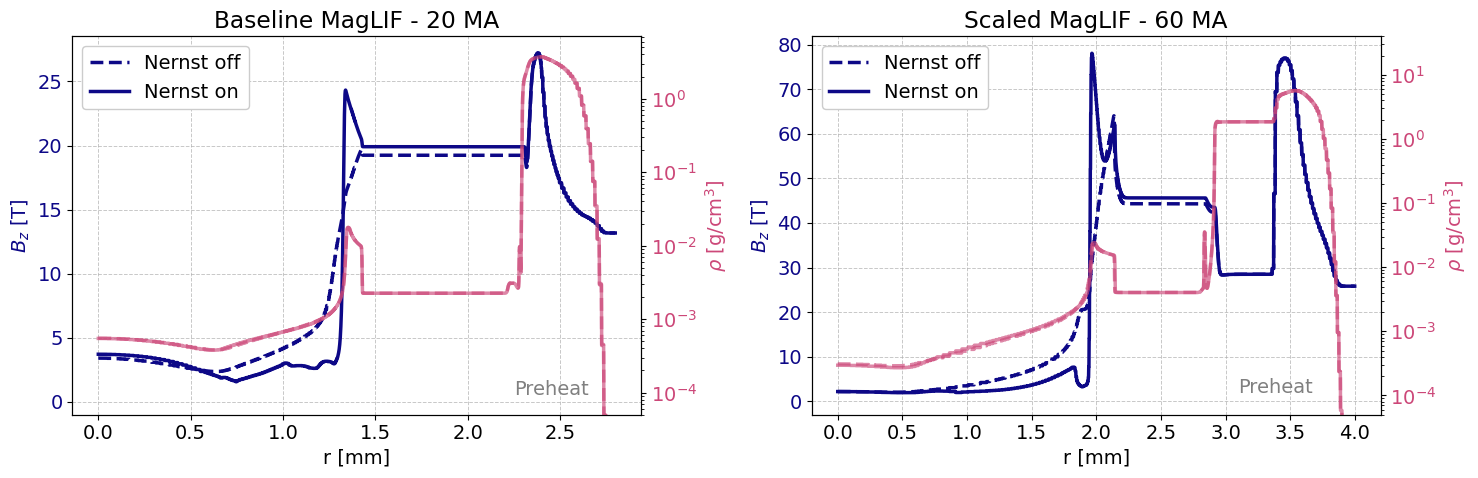}
    \caption{Axial magnetic field $B_z$ and density $\rho$ profiles at the end of preheat (90 ns) from 1D MagLIF FLASH simulations for 20 MA current drive (left) and 60 MA current drive (right).}
    \label{fig:maglif profiles preheat}
\end{figure}

For further insight into the preheat phase, profiles for axial magnetic field and density at the end of the laser pulse are presented in Fig. \ref{fig:maglif profiles preheat}.
The qualitative features of the propagation are well described by the theoretical model discussed in this work. 
Notably, the laser compresses the fuel into an outer shelf that surrounds a low-density inner core, the former being delimited by the leading shock and a trailing ablation front.  
The magnetic field profile follows a similar arrangement. 
It is compressed by the leading shock in a case-dependent manner--showcased by the weak compression observed in the baseline configuration, in contrast to the stronger shock-driven compression in the scaled case.
The magnetic field diffuses into the inner core, forming a boundary layer placed at the ablation front.
The Nernst term opposes this diffusion by efficiently advecting magnetic field from the core towards the shelf, resulting in a thinner boundary layer. 
Although not shown in Fig. \ref{fig:maglif profiles preheat} for clarity, the characteristic electron Hall values in the outermost layers of the core are approximately 1 and 10 in the baseline and scaled cases, respectively. 
These values are in agreement with those predicted by Eq. \eqref{eq:ehall numerical} using Table \ref{tab:maglif input paramteres} as input conditions, reading 1.94 and 5.67.
We report however that despite Nernst convection, the central part of the core is not demagnetized by this time.
This is in contrast to the preheat simulation results shown in Fig. \ref{fig:preheat profiles magnetized}, which predicted complete central demagnetization by the end of the laser pulse.
A potential explanation for this discrepancy lies in the ion-electron equilibration time being considerable within the laser hot spot.
Consequently, electrons remain hotter than the ions in the central part of the core, favoring larger electron Hall values, which reduces Nernst convection in this region.

\begin{figure}[h]
    \centering
    \includegraphics[width=0.85\linewidth]{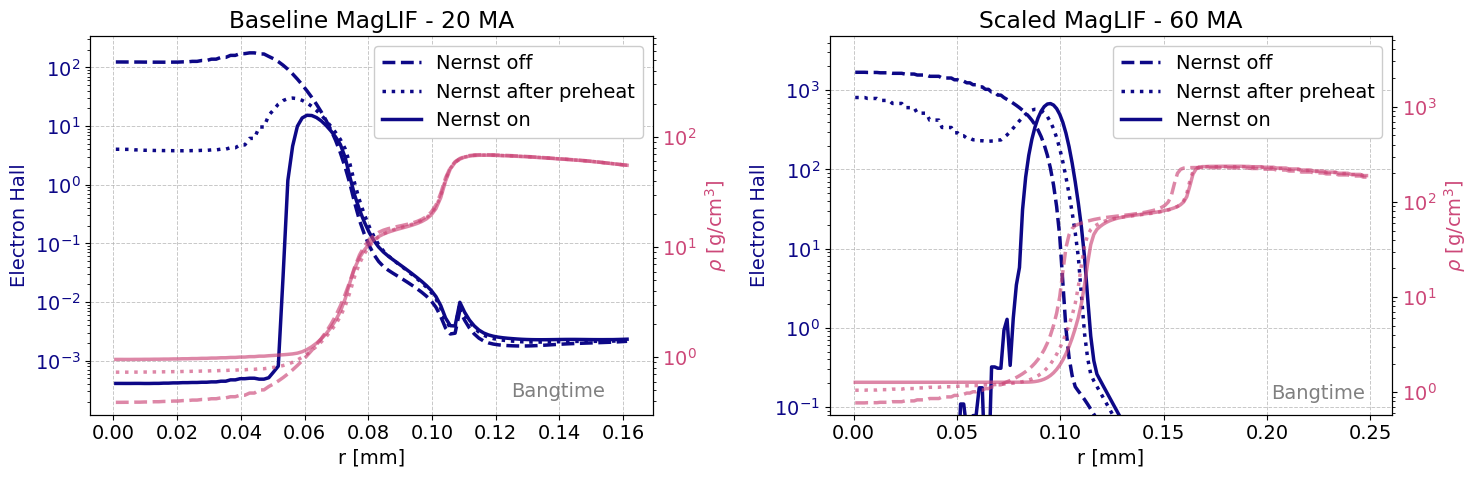}
    \caption{Electron Hall parameter and density $\rho$ profiles from 1D MagLIF FLASH simulations for 20 MA current drive (left) and 60 MA current drive (right). The selected times correspond to bangtime (152 ns for 20 MA and 150 ns for 60 MA). The oscillations in the magnetic field profile in the bottom right panel are numerical and associated to low $B_z$ field values.}
    \label{fig:maglif profiles stagnation}
\end{figure}

The dynamics during preheat influence the hot spot conditions established later in the implosion, as illustrated by the profiles of density and electron Hall parameter at bangtime in Fig. \ref{fig:maglif profiles stagnation}.
We first note that, at this time, the interface between the fuel and liner (defined as the position where the beryllium mass concentration reaches 0.5) is located at 0.077 mm and 0.113 mm for the baseline and scaled cases, respectively, in the Nernst-on runs.
We can see that a denser, colder layer of DT fuel surrounds the hot core at bangtime.
This layer is reminiscent of the outer fuel shelf formed earlier during the preheat phase, which eventually impacts the liner and implodes in unison.
When Nernst is accounted for, the region exhibiting the highest levels of magnetization is localized within the outermost hot spot layer adjacent to this colder fuel layer.  
As showcased by the `Nernst after preheat' runs, including Nernst during the preheat phase is crucial for accurately capturing this structure later on at stagnation.
Although it could be argued that only the electron Hall values in the outer layers are relevant for thermal insulation of the hot spot, the Nernst-off runs exhibit overall larger electron Hall values, affecting the integrated performance of the configuration.

\begin{table}[h]
    \centering
    \begin{tabular}{c|ccc}
          & 1D FLASH - Nernst off  & 1D FLASH - Nernst on & 2D HYDRA Ref. \cite{Ruiz_2023_Iscaling} - Nernst on \\ 
         \hline
         \hline
         Alpha off & 15.7 MJ & 10.3 MJ & 5.4 MJ\\
         Alpha on & 43.3 MJ & 62.1 MJ  & 60.2  MJ
    \end{tabular}
    \caption{Fusion yield for scaled 60 MA MagLIF simulations. Alpha-off runs do not consider the alpha energy deposition back into the fuel, while alpha-on runs propagate and deposit alpha energy according to a diffusion model. In the 1D FLASH simulations, the yield is obtained by multiplying the output yield per unit length by the corresponding imploding length of 1.78 cm. The 2D HYDRA data is obtained from Figure 19 in Ref \cite{Ruiz_2023_Iscaling}.}
    \label{tab:yields}
\end{table}

We finally discuss fusion performance for the scaled 60 MA MagLIF configuration, listed in Table \ref{tab:yields}. 
The ‘alpha off’ label refers to simulations in which the energy deposition from alpha particles into the fuel is neglected, while ‘alpha on’ accounts for this feedback using a diffusion model for the propagation of alpha energy.
The simulations including Nernst show good agreement with the 2D HYDRA results reported in the original current-scaling study \cite{Ruiz_2023_Iscaling}. 
The alpha-off/Nernst-off run performs better than its Nernst-on equivalent, a direct consequence of the weaker thermal insulation occurring in the latter due to a more degraded field compression. 
The trend is reversed for the alpha-on runs, which undergo significant fuel self-heating.
This behavior arises because the fuel mass partition achieved at bangtime qualitative resembles that of the hot spot conditions in high-gain MagLIF configurations \cite{Slutz2012HighGain}, where an outer cryogenic DT layer encloses a low-density vapor region. 
These `ice-burn' configurations rely on igniting a hotter central fuel region, initiating a burn wave that propagates into adjacent colder fuel layers.
Due to preheat propagation, a similar situation is achieved for the `gas-burn' configurations discussed here (which depart from a homogeneous fuel density profile), albeit with substantially less mass inventory in the cold shelf.
Burn propagation is sensitive to the magnetic field as the diffusion of the energy of the alpha particles is reduced with magnetization \cite{liberman1984distribution}. 
The alpha-on/Nernst-off case leads to excessive hot spot magnetization, presumably hindering burn propagation into the colder shelf, resulting in lower yield.
Slutz \textit{et al.} \cite{Slutz2016Scaling} discussed a similar mass-partitioning effect in gas-burn configurations arising during preheat, and identified an optimal magnetic field of 10 T at 60 MA, which allowed some of the cold fuel surrounding the hot core to be burned.
This highlights the importance of accurately modeling the preheat phase to capture burn sensitivity to the initial magnetic field value.

\section{Conclusions}
\label{sec:conclusions}

In this work, we present a mathematical model to investigate preheat propagation in a magnetized plasma.
The model formally assumes that the time scale for pressure disturbances to propagate is much shorter than that for thermal conduction, a condition quantified by the smallness of the dimensionless parameter $\epsilon$ defined in Eq. \eqref{eq:epsilon}.
This is the regime that characterizes the physics of preheat in MagLIF configurations.
The resulting fluid expansion exhibits two distinct structures: an inner, hot, low density core surrounded by an outer, cold, denser shelf bounded by a leading strong shock. 
This structure determines the magnetic field distribution, which is compressed by the shock, accumulates in the outer shelf, and diffuses into the inner core.
We have assessed the implications of this mass and magnetic field partitioning that emerges during preheat on the thermal insulation, flux compression, and performance of integrated MagLIF simulations. 

We have first extended the self-similar analysis of preheat propagation in a gas made in Ref. \cite{kurdyumov2003heat} by introducing a normalization more appropriate for cases where strong pressures arise from energy deposition.
As a result, we have derived numerical expressions for the characteristic pressures, temperatures, and densities in each region, expressed in terms of the input parameters [Eqs. \eqref{eq:p0 numerical}--\eqref{eq:temp core numerical}].
Additionally, we have characterized the interface between the core and the shelf as an ablation front through which mass is ablated from the shelf to replenish the core.
We have also identified self-similar expansion solutions corresponding to power deposition laws that follow a power-law dependence in time, as well as to the post-deposition phase following source shutoff.

The analysis of the dynamics of a background magnetic field during preheat represents a novel contribution of this work. 
While we have not derived semi-analytical solutions for the field profile, since magnetic diffusion breaks self-similarity, we have qualitatively described the field dynamics and obtained numerical expressions for the characteristic magnetic Reynolds number ($\rem$) and expected electron Hall parameters [Eqs. \eqref{eq:rem numerical} and \eqref{eq:ehall numerical}].
FLASH simulations of preheat scenarios under typical MagLIF conditions confirm the adequacy of these expressions.
These simulations assess that the strength of the field compression exerted by the shock is sensitive to the input conditions, while its subsequent evolution in the outer shelf is nearly ideal, characterized by a large $\rem$.
The magnetic field then penetrates into the inner core primarily through diffusion, while Nernst convection acts in opposition, resulting in a magnetic boundary layer forming at the ablation front.
As a consequence, the region where significant magnetization occurs, characterized by finite electron Hall values, is localized adjacent the ablation front. 

This model provides a qualitative framework for describing preheat physics in FLASH simulations of MagLIF configurations corresponding to the 20 MA and 60 MA current-drive cases defined in the current-scale study of Ref. \cite{Ruiz_2023_Iscaling}.
In particular, it predicts the partitioning of mass and magnetic field observed within the structures that develop during the expansion, although it tends to overestimate the localization of magnetization effects due to the assumption of instantaneous ion-electron thermal equilibration.
This stratification has implications in the magnetic flux conservation and fusion performance of the subsequent implosion.
Significant amount of flux (approximately thirty percent) is lost from the fuel immediately after the blast wave reaches the liner in both configurations.
The denser fuel shelf implodes coherently with the liner once it comes in contact with it, leading to a hot spot configuration at bangtime where a central hot core is surrounded by a colder fuel layer. 
Magnetization effects at this stage of the implosion are well localized in the outermost layers of the core, which is otherwise the critical barrier for thermal insulation.
However, performance of the igniting 60 MA configuration is adversely affected by excessive magnetization levels, as they impede the ability of the burn wave to propagate into the colder fuel layer.
This tradeoff highlights the importance of resolving the described flow structures and scales that arise during the preheat phase in order to appropriately calculate the fusion yield of MagLIF configurations.

This mathematical model can also provide useful guidelines in the design of MagLIF targets. 
The fuel mass stratification is governed by the parameter $\epsilon$, which shows a strong dependence on both the laser pulse duration and the initial fuel density [Eq. \eqref{eq:eps numerical}]. 
Increasing this parameter reduces the shelf-to-core density ratio. 
Separately, the self-similar solutions derived shall be understood as a fluid state attained asymptotically in time. 
Reducing the time that it takes for the preheat-driven expansion to reach the inner liner surface (relative to the laser deposition time) would prevent this state to completely establish.
This could be controlled for example by modifying the characteristic expansion velocity $r_h/t_d$, with $r_h$ given by Eq. \eqref{eq:rh numerical}.
Finally, the core demagnetization could be alleviated by enhancing more magnetic field diffusion into it. 
This could be achieved by lowering the magnetic Reynolds number [Eq. \eqref{eq:rem numerical}], which presents relatively strong dependence on the three input parameters considered in the model. 

\appendix
\section{Verification test: Nernst thermal wave}
\label{sec:nernst thermal wave}

We present FLASH simulations of subsonic thermal waves evolving in a magnetized plasma, a problem previously studied theoretically by Garcia-Rubio et al. \cite{garcia2017mass}, among others.
This problem provides a useful benchmark for building confidence in the simulation of the hot spot dynamics of magnetized ICF concepts such as MagLIF. 
It considers the evolution of a semi-infinite region of hot, magnetized deuterium plasma in contact with a cold, unmagnetized plasma (wall) of the same material in mechanical equilibrium.
Particularly, the analysis in Ref. \cite{garcia2017mass} accounts for the ablation of the cold material driven by heat flux, a key physical process in hot spot evolution. 

The resulting motion is self-similar with respect to the variable $x/\sqrt{t}$.
The thermal diffusivity of the hot plasma neglecting magnetization effects ($\kappa _0$) is used to nondimensionalize this variable. 
The two dimensionless parameters that govern the structure of this problem are the magnetic Lewis number $\lem$, defined as the ratio between the thermal and the magnetic diffusivities [Eq. (15) in Ref. \cite{garcia2017mass}], and the electron Hall parameter in the hot plasma $x_{e0}$.
The analysis assumes negligible magnetic pressure effects, which imposes the constraint $\lem \gg x_{e0}^2$. 
For the interested reader, such effects were considered in a subsequent publication \cite{garcia2018pressure}.

The hot plasma heats up the cold material, leading to its ablation. 
Layers of cold plasma thereby enter the hot spot, forming a thermal wave structure whose thickness, in self-similar variables, scales inversely with the magnetization level as $x_{e0}^{-7/10}$.
Larger electron thermal parameters result in reduced thermal diffusivity, a thinner wave structure, and decreased mass ablation. 
The magnetic field, on its part, is subject to flow advection by the ablated plasma, Nernst convection towards the wall, and diffusion. 
The magnetic Lewis number assesses the relative importance of the latter with respect to the convective terms. 
When $\lem$ is large, diffusion is confined to thin boundary layers within the wave structure. 
When the plasma is unmagnetized, Nernst dominates and the magnetic field is piled up against the ablation front, presenting a local peak adjacent to it. 
Magnetization effects inhibit Nernst transport, which is overcome by flow convection when the local electron Hall value exceeds 4.37 in this problem. 
The B-field structure exhibits two peaks for sufficiently large values of $x_{e0}$: one due to Nernst convection in the weakly magnetized, colder plasma, and a second, deeper within the hot spot, due to field compression by the flow.

\begin{figure}
    \centering
    \includegraphics[width=0.85\linewidth]{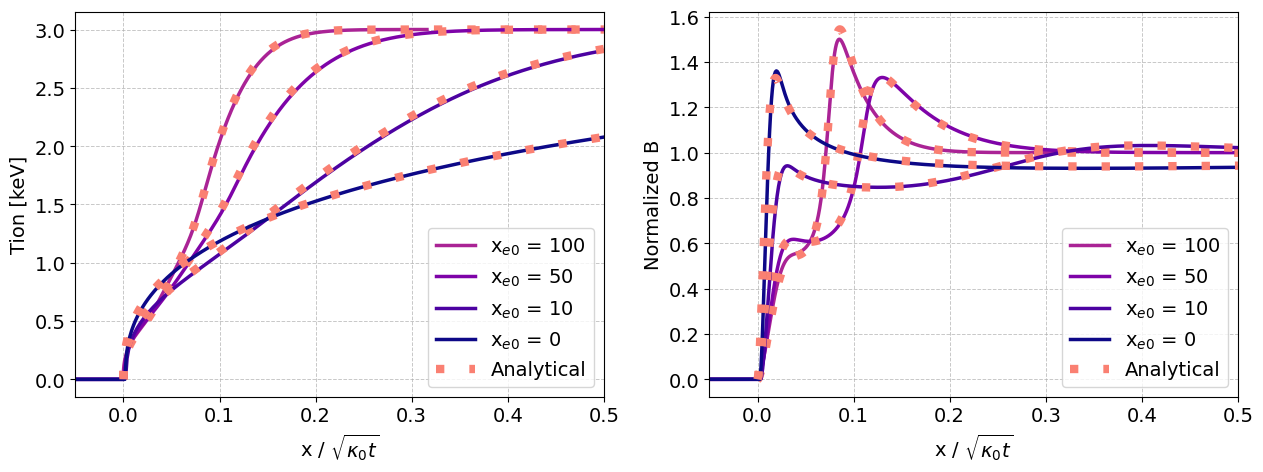}
    \caption{Ion temperature and magnetic field ($B$) self-similar profiles of FLASH simulations of subsonic Nernst thermal waves for different electron Hall parameters $x_{e0}$ and a magnetic Lewis number $\lem = 10^5$. In each case, the magnetic field has been normalized with its value downstream. The dotted lines correspond to the semi-analytical profiles from Fig. 5 in Ref. \cite{garcia2017mass}.}
    \label{fig:nernst}
\end{figure}

Temperature and magnetic field (B) profiles for a large $\lem = 10^5$ and different $x_{e0}$ from FLASH simulations are compared to the theoretical analysis in Fig. \ref{fig:nernst}.
These conditions were attained with an initial hot plasma of 50 mg/cm$^3$ density and 3 keV temperature, and magnetic field intensities spanning 0.001, 1.055, 5.275, and 10.55 MG.
The 
The time step of the profiles plotted corresponds to 20 ns. 
An artificially large density jump of $10^5$ was imposed to ensure that the ablation front remained fixed at $x = 0$ throughout the simulation.
The FLASH simulations presented perfectly capture the phenomenology described and accurately reproduce the analytical profiles. 
The slight underprediction of the B-field peak in the case $x_{e0} = 100$ is attributed to magnetic pressure effects, as the inequality $\lem \gg x_{e0}^2$ is not strictly satisfied in this case.

\bibliographystyle{unsrt}
\bibliography{refs_heat}

\end{document}